\documentclass[journal]{IEEEtran}

\usepackage{cite}
\usepackage{siunitx}
\usepackage{amsmath}
\usepackage{multirow}
\usepackage{amsmath}
\usepackage{amssymb}
\usepackage{booktabs} 
\usepackage{url}
\usepackage{multirow}
\usepackage{graphicx}
\usepackage{enumerate}
\usepackage{bm}
\usepackage{color}
\usepackage{gensymb,textcomp}
\usepackage{tabularx}
\usepackage{array}

\ifCLASSOPTIONcompsoc
  \usepackage[caption=false,font=normalsize,labelfont=sf,textfont=sf]{subfig}
\else
  \usepackage[caption=false,font=footnotesize]{subfig}
\fi

\ifCLASSINFOpdf
\else
\fi

\graphicspath{
    {./images/}{./images/proj/}
}

\newcommand{\dg}{\textdegree}
\newcommand{\tabincell}[2]{\begin{tabular}{@{}#1@{}}#2\end{tabular}}

\newcolumntype{L}[1]{>{\raggedright\arraybackslash}m{#1}}
\newcolumntype{C}[1]{>{\centering\arraybackslash}m{#1}}
\newcolumntype{R}[1]{>{\raggedleft\arraybackslash}m{#1}}

\begin{document}
%
\title{State-of-the-art in 360{\dg} Video/Image Processing: Perception, Assessment and Compression}
%
%
%

\author{Chen Li,~\IEEEmembership{Student Member,~IEEE,}  Mai Xu,~\IEEEmembership{Senior Member,~IEEE,} Shanyi Zhang, Patrick Le Callet,~\IEEEmembership{Fellow,~IEEE}
\thanks{Manuscript received April 16, 2019.}}

%
%

\markboth{Journal of \LaTeX\ Class Files,~Vol.~14, No.~8, August~2015}%
{Shell \MakeLowercase{\textit{et al.}}: Bare Demo of IEEEtran.cls for IEEE Journals}
%



\maketitle

\begin{abstract}
Nowadays, 360{\dg} video/image has been increasingly popular and drawn great attention. The spherical viewing range of 360{\dg} video/image accounts for huge data, which pose the challenges to 360{\dg} video/image processing in solving the bottleneck of storage, transmission, etc. Accordingly, the recent years have witnessed the explosive emergence of works on 360{\dg} video/image processing. In this paper, we review the state-of-the-art works on 360{\dg} video/image processing from the aspects of perception, assessment and compression. First, this paper reviews both datasets and visual attention modelling approaches for 360{\dg} video/image. Second, we survey the related works on both subjective and objective visual quality assessment (VQA) of 360{\dg} video/image. Third, we overview the compression approaches for 360{\dg} video/image, which either utilize the spherical characteristics or visual attention models. Finally, we summarize this overview paper and outlook the future research trends on 360{\dg} video/image processing.
\end{abstract}


%
\IEEEpeerreviewmaketitle

\section{Introduction}
%
%
%
%
\IEEEPARstart{3}{60{\dg}} video/image, also known as panoramic, spherical or omnidirectional video/image, is a new multimedia type that provides immersive experience. The content of 360{\dg} video/image is on the sphere that covers the whole $360\times180\degree$ viewing range. In other words, 360{\dg} video/image surrounds the viewer seamlessly and occupies the entire vision of the viewer, which is different from traditional 2-dimensional (2D) video/image that only covers a limited plane. Recent years have witnessed the rapid development of virtual reality (VR) technology. As an essential type of VR content, 360{\dg} video/image has been flooding into our daily life and drawing great attention. With the commercial head mount displays (HMDs) available recently, the viewer is allowed to freely make the viewport focus on the desired content via head movement (HM), just like humans do in real world. In this way, the immersive and even interactive experiences are technically achieved.
Meanwhile, new challenges have been raised to 360{\dg} video/image processing. To cover the whole $360\times180\degree$ viewing range with high fidelity, the resolution of 360{\dg} video/image is extraordinarily high. Moreover, for 360{\dg} video, the frame rate should also be high to avoid motion sickness of viewers \cite{chen2018recent}. Consequently, heavy burdens are laid on the storage and transmission of 360{\dg} video/image.

To relieve the storage and transmission burdens, compression is in urgent need for saving bitrates of 360{\dg} video/image. In the past decades, many video/image compression standards have been developed for traditional 2D video/image by the organizations of International Telecommunication Union (ITU), Joint Photographic Experts Group (JPEG), Moving Picture Experts Group (MPEG), etc. Although 360{\dg} images and video sequences are projected from sphere to 2D planes to be stored, processed and transmitted \cite{chen2018recent}, these 2D standards do not fit 360{\dg} video/image well due to the spherical characteristics. Therefore, projects for improving compression efficiency on 360{\dg} video/image were started, such as MPEG-I \cite{wien2019standardization} and JPEG 360 \cite{jpeg2017call}, and much effort have been made in 360{\dg} video/image compression \cite{wien2019standardization}.
To measure the compression performance, visual quality assessment (VQA) is needed to evaluate the quality degradation caused by compression. However, due to the spherical characteristics and the existence of sphere-to-plane projection, subjective VQA recommendations \cite{itur2007methodology,itut2008subjective,itur2012methodology} and objective VQA approaches \cite{seshadrinathan2010study} for 2D video/image are not appropriate for 360{\dg} video/image. Under this circumstance, there have emerged several works for VQA on 360{\dg} video/image, from the aspects of effectively collecting subjective quality data \cite{upenik2016testbed,xu2017subjective} and modelling the visual quality \cite{chen2018recent}.
Moreover, the unique mechanism of viewing 360{\dg} video/image through the viewport in an HMD accounts for two facts: (1) The quality degradation in the viewport is more noticeable in 360{\dg} video/image, since the viewer focus on the viewport, which is a small part of the whole 360{\dg} video/image. (2) There is massive redundancy in the encoded bits of 360{\dg} video/image, since the giant region outside the viewport is invisible to the viewer. Inspired by these facts, consideration of human perception may benefit VQA and compression on 360{\dg} video/image. Thus, there are also many works concentrating on visual attention modelling for 360{\dg} video/image to predict human perception \cite{gutierrez2018toolbox}, and even the grand challenges were held \cite{rai2017salient360,gutierrez2018salient360}.

\begin{table*}[t]%
\centering%
\caption{Summary of the existing 360{\dg} video/image datasets with the visual attention data of subjects.}\label{tab:sal:ds}%
\resizebox{\textwidth}{!}{%
\begin{tabular}{|c|c|c|m{8.5em}|m{6em}|c|c|m{30em}|}%
\hline
Dataset & Image/Video & Subjects & \multicolumn{1}{c|}{Dataset Size} & \multicolumn{1}{c|}{Resolution} & Duration * & HM/EM & \multicolumn{1}{c|}{Description} \\
\hline
Abreu \textit{et al.} \cite{de2017look} & Image & 32 & 21, indoor and outdoor images & 4096$\times$2048 pixels & 10 or 20 s & HM & HMD: Oculus Rift DK2. The subjects were equally divided into two groups. For one group, the 360{\dg} images were viewed for 10 seconds, and for the other group, the 360{\dg} images were viewed for 20 seconds. \\
\hline
Bao \textit{et al.} \cite{bao2016shooting} & Video & \tabincell{c}{153\\Ages: 20-50} & \multicolumn{1}{c|}{16 from 3 categories} & Full HD to 4K, mostly 4K & --- & HM & HMD: Oculus DK2. 35 subjects viewed all 16 sequences in order, and the remained 118 watched 3-5 randomly selected sequences. \\
\hline
Ozcinar \textit{et al.} \cite{ozcinar2018visual} & Video & 17 & \multicolumn{1}{c|}{6} & \multicolumn{1}{c|}{4K to 8K} & 10 s & HM & HMD: Oculus Rift CV1. \\
\hline
Corbillon \textit{et al.} \cite{corbillon2017360} & Video & \tabincell{c}{59\\Ages: 6-62} & \multicolumn{1}{c|}{7} & 3840$\times$2048 pixels & 70 s & HM & HMD: Razer OSVR HDK2. \\
\hline
Lo \textit{et al.} \cite{lo2017360} & Video & 50 & \multicolumn{1}{c|}{10 from 3 categories} & \multicolumn{1}{c|}{4K} & 1 min & HM & HMD: Oculus Rift DK2. \\
\hline
Wu \textit{et al.} \cite{wu2017dataset} & Video & \tabincell{c}{48\\Ages: 20-26} & \multicolumn{1}{c|}{18 from 5 categories} & Full HD to 4K, mostly 2K & $\approx$ 3-11 min & HM & HMD: HTC Vive. Half of the sequences were freely viewed by subjects. Before viewing the other half, subjects were instructed with questions. \\
\hline
AVTrack360 \cite{fremerey2018avtrack360} & Video & \tabincell{c}{48\\Ages: 18-65} & \multicolumn{1}{c|}{20} & \multicolumn{1}{c|}{4K} & 30 s & HM & HMD: HTC Vive. Although free viewing was allowed, subjects were asked to fill in the simulator sickness questionnaire during the test session. \\
\hline
VR-HM48 \cite{xu2017subjective} & Video & 40 & \multicolumn{1}{c|}{48 from 8 categories} & \multicolumn{1}{c|}{3K to 8K} & 20-60 s & HM & HMD: HTC Vive. No intercutting. \\
\hline
Wild-360 \cite{cheng2018cube} & Video & 30 & \multicolumn{1}{c|}{85 from 3 categories} & \multicolumn{1}{c|}{---} & --- & HM-like & The sequences in this dataset were not viewed in HMD. Subjects viewed the sequences under projection in a typical 2D monitor and labelled the viewport positions with mouse. As the sequences were not mapped onto the sphere, the data collected in this dataset are not real HM data. \\
\hline
Sitzmann \textit{et al.} \cite{sitzmann2018saliency} & Image & \tabincell{c}{169\\Ages: 17-59} & \multicolumn{1}{c|}{22} & \multicolumn{1}{c|}{---} & 30 s & HM+EM & HMD: Oculus Rift DK2. Eye-tracker: pupil-labs stereoscopic eye-tracker. There were three experiment conditions: the VR stand condition, the VR seated condition and the desktop condition. In the desktop condition, the data were collected using Tobii EyeX eye-tracker, with mouse-controlled desktop panorama viewers in a typical 2D monitor,  \\
\hline
\multirow{2}[5]{*}{Salient360} & Image \cite{rai2017dataset,gutierrez2018toolbox} & \tabincell{c}{63\\Ages: 19-52} & 98 (60 is released) from 5 categories & 5376$\times$2688 to 18332$\times$9166 pixels & 25 s & HM+EM & HMD: Oculus Rift DK2. Eye-tracker: Sensomotoric Instruments (SMI) eye-tracker. Each subject viewed a subset of 60 images, so each image was viewed by 40-42 subjects. \\
\cline{2-8}  & Video \cite{david2018dataset} & \tabincell{c}{57\\Ages: 19-44} & 19, categorized by 3 groups of labels & 3840$\times$1920 pixels & 20 s & HM+EM & HMD: HTC Vive. Eye-tracker: SMI eye-tracker. No intercutting. \\
\hline
PVS-HM \cite{xu2018predicting} & Video & \tabincell{c}{58\\Ages: 18-36} & \multicolumn{1}{c|}{76 from 8 categories} & \multicolumn{1}{c|}{3K to 8K} & 10-80 s & HM+EM & HMD: HTC Vive. Eye-tracker: aGlass DKI. \\
\hline
Zhang \textit{et al.} \cite{zhang2018saliency} & Video & \tabincell{c}{27\\Ages: 20-24} & 104 from 5 sports catogories & \multicolumn{1}{c|}{---} & 20-60 s & HM+EM & HMD: HTC Vive. Eye-tracker: aGlass DKI. The sequences are from the sports-360 dataset \cite{hu2017deep}. Each sequence was viewed by at least 20 subjects. \\
\hline
VR-EyeTracking \cite{xu2018gaze} & Video & \tabincell{c}{45\\Ages: 20-24} & 208 from more than 6 categories & \multicolumn{1}{c|}{$>$ 4K} & 20-60 s & HM+EM & HMD: HTC Vive. Eye-tracker: aGlass DKI. The sequences were divided into 6 groups, and subjects viewed one group each time. Each sequence was viewed by at least 31 subjects. \\
\hline
VQA-ODV \cite{li2018bridge} & Video & \tabincell{c}{221\\Ages: 19-35} & 60 from more than 4 categories (the other 540 are with the same content as the 60 sequences) & 3840$\times$1920 to 7680$\times$3840 pixels & 10-23 s & HM+EM & HMD: HTC Vive. Eye-tracker: aGlass DKI. No intercutting. The task for subjects was assessing visual quality of the sequences instead of free viewing. The sequences were equally divided into 10 groups, and each subject only viewed 1 group of sequences (60 sequences in total including 6 reference sequences). Each sequence was viewed by at least 20 subjects. \\
\hline
\multicolumn{8}{l}{* For videos, duration represents the temporal length of the videos themselves; For images, duration represents the time each subject view each image in the experiment.}\\
\end{tabular}%
}%
\end{table*}%
Even though plenty of works targeting at 360{\dg} video/image processing have dramatically emerged in recent years, to the best of our knowledge, there lacks a survey that reviews on these works and provides summary and outlook. In this paper, we survey works on 360{\dg} video/image processing from the aspects of visual attention modelling, VQA and compression, and we also reveal the relationships among these aspects. For visual attention modelling, we review both datasets and approaches. For VQA, subjective VQA methods for 360{\dg} video/image are first surveyed, which also provide datasets with subjective quality scores. Then, we focus on objective VQA approaches, which aim at being consistent with the subjective quality scores on 360{\dg} video/image. For compression, approaches that either incorporate the spherical characteristics or perception of 360{\dg} video/image are reviewed. At last, we summarize our paper and outlook the future research trends on 360{\dg} video/image processing.

\section{Visual attention modelling}\label{sec:sal}
Different from traditional 2D video/image, 360{\dg} video/image can be perceived in the range of $360\times180\degree$. 
Humans can select the viewport to focus on the attractive content of 360{\dg} video/image through their HM within a sphere, while the eye movement (EM) determines which region can be captured at high resolution (\textit{i.e.}, fovea) within the viewport.
In particular, the HM refers to the locations of the subjects' viewports, while the EM reflects region-of-interest (RoI) inside the viewports.  
Therefore, the main difference between the perception models of 360{\dg} video/image and 2D video/image is the visual attention mechanism.  
In other words, it is necessary to predict both HM and EM for predicting visual attention on 360{\dg} video/image as the perception model. 
In this section, we review the datasets and approaches for visual attention modelling on 360{\dg} video/image.

\subsection{Datasets}\label{sec:sal:ds}
For establishing a visual attention dataset for 360{\dg} video/image, the HM and EM data can be collected in the VR environment (except Wild-360 \cite{cheng2018cube}), \textit{i.e.}, subjects view 360{\dg} content by wearing the HMD. 
Specifically, the posture data of HMD can be captured via the software development kit (SDK). Based on the posture data, the HM data of subjects can be calculated and recorded.
In addition, the eye-tracker, which can be embedded into the HMD, can be used to track the pupil and corneal reflections for capturing the EM data.

Recently, there have emerged many datasets for visual attention modelling on 360{\dg} video/image, which contain the HM data and even the EM data of subjects. Table \ref{tab:sal:ds} summarizes these datasets.
Note that all the datasets listed in Table \ref{tab:sal:ds} are public and can be downloaded online.
The datasets of Table \ref{tab:sal:ds} enable the analysis on human attention, when subjects viewing 360{\dg} video/image. Additionally, these datasets also boost the data-driven approaches for modelling human attention. 
Some of widely used datasets are discussed in more details as follows.

\textbf{Abreu \textit{et al.}} \cite{de2017look} is one of the earliest attention dataset for 360{\dg} image, which was built in 2017. It contains 21 360{\dg} images, including indoor and outdoor scenes. All the images are with resolution of $4096\times2048$ pixels. In total, there are 32 subjects involved in viewing the 360{\dg} images. In the experiment, the HMD of Oculus Rift DK2 was used to collect the HM data of subjects. The subjects were divided into two groups, both with 16 subjects. For one group, the 360{\dg} images were viewed for 10 seconds, and for the other group, the 360{\dg} images were viewed for 20 seconds. Before the test session for data collection, there is a training session in their experiment to make subjects be familiar with the HMD and 360{\dg} images. Then, they were allowed to freely view the images in the test session, and the posture data of HMD were captured.
Consequently, the HM data of 32 subjects were obtained for the dataset. 

\textbf{Salient360} datasets contain both the HM and EM data for 360{\dg} image \cite{rai2017dataset,gutierrez2018toolbox} and video \cite{david2018dataset}. They are the popular datasets, which have been widely used as the training sets or benchmarks in many recent visual attention models \cite{startsev2018360,lebreton2018gbvs360,luz2017saliency,battisti2018feature,zhu2018prediction,fang2018novel,ling2018saliency,assens2018scanpath,monroy2018salnet360,chao2018salgan360,suzuki2018saliency,lebreton2018v}. In  \cite{rai2017dataset} and \cite{gutierrez2018toolbox}, 98 360{\dg} images are included from 5 categories: small indoor rooms, grand halls, natural landscapes, cityscapes and people. Among them, 60 are currently released. The resolution ranges from $5376\times2688$ pixels to $18332\times9166$ pixels. Then, 63 subjects were asked to view the 360{\dg} images for 25 seconds through the HMD, and each subject freely viewed a subset of 60 images. As such, each image was viewed by 40-42 subjects. In the Salient360 dataset of 360{\dg} video \cite{david2018dataset}, 19 video sequences with resolution of $3840\times1920$ pixels are included, categorized by 3 groups of labels. Each sequence is with duration of 20 seconds. In total of 57 subjects freely viewed all sequences in the experiment. 
It is worth mentioning that subjects were seated in a swivel chair, when they viewed the 360{\dg} videos/images.
For both datasets, an HMD-embedding eye-tracker from SMI is utilized, so that not only the HM data but also the EM data of subjects were collected for the Salient360 datasets. 

\textbf{VR-EyeTracking} \cite{xu2018gaze} is a large-scale dataset for 360{\dg} video, containing both the HM and EM data of subjects. In total, 208 video sequences are included, with diverse content of indoor scene, outdoor activities, music shows, sports games, documentation, short movies, etc. For each sequence, the resolution is at least 4K, and the duration ranges from 20 to 60 seconds. The HM and EM data were captured by the HMD of HTC Vive embedded with the eye-tracker of aGlass. In the experiment, 45 subjects were recruited to view the video sequences. The sequences were divided into 6 groups, and subjects freely viewed one group each time. The re-calibration of the eye-tracker was implemented in interval between two groups.
Each sequence was viewed by at least 31 subjects. 
Finally, the VR-EyeTracking dataset was established.

Based on the existing datasets, there are several works analyzing human attention on 360{\dg} video/image from the following aspects. 
Note that the dataset analysis can be further used to facilitate visual attention modelling.

\begin{enumerate}
  \item \textbf{Consistency among subjects}. 
  It is essential to analyze the consistency of visual attention among different subjects, when viewing 360{\dg} video/image.
  To this end, human attention is modelled in the form of 2D saliency maps, which are the heat maps of HM/EM fixations, projected from the sphere of 360{\dg} video/image to the 2D plane. 
  Then, the similarity of saliency maps generated from different subjects indicates the consistency of visual attention among subjects.
  In \cite{sitzmann2018saliency}, Sitzmann \textit{et al.} compared the saliency maps of 360{\dg} images between each subject and the remaining subjects through the receiver operating characteristic (ROC) curve. 
  The fast convergence of the ROC curve to the maximum rate of 1 indicates high consistency among subjects.
  Quantitatively, 70\% of EM fixations fall within the 20\% most salient regions in the analysis of \cite{sitzmann2018saliency}. 
  For 360{\dg} video, Xu \textit{et al.} \cite{xu2018predicting} proposed calculating Pearson's linear correlation coefficient (PLCC) of the HM saliency maps between two groups of subjects, each of which includes half of the subjects. The PLCC result was reported to be 0.83 \cite{xu2018predicting}, which is significantly higher than the random baseline. In summary, the consistency of human attention on 360{\dg} video/image is high across subjects.
  
  \item \textbf{Equator and center bias}. 
  When viewing 2D video and image, humans prefer to looking at the center of video/image.
  In other words, there exists the center bias for the EM data on 2D video/image.
  Similarly, the statistical bias also holds for the HM/EM fixations on 360{\dg} video/image.  
  For 360{\dg} image, it is found in  \cite{sitzmann2018saliency} and \cite{lebreton2018gbvs360} that the subjects tend to view the regions near the equator more frequently, called the equator bias.
  The equator bias also exists for human attention on 360{\dg} video. 
  In addition, it has been investigated in the latest work of \cite{xu2018predicting} and \cite{david2018dataset} that subjects watch the front region of 360{\dg} video much more frequent. 
  Therefore, human attention on 360{\dg} video is biased toward both the equator and the front region, which is different from 360{\dg} image.
  The front bias and the equator bias can be used as prior knowledge in the visual attention models of  360{\dg} video/image \cite{startsev2018360,lebreton2018gbvs360,luz2017saliency,battisti2018feature,zhu2018prediction,fang2018novel,ling2018saliency,suzuki2018saliency,lebreton2018v,xu2018predicting} to improve their prediction accuracy.
 
  \item \textbf{Impact of content on attention.} It has been illustrated in \cite{xu2018assessing} that in addition to the statistical bias, human attention are highly correlated with the content of 360{\dg} video. For example, the salient objects of 360{\dg} video have potential in attracting human attention.
  This was quantitatively verified in \cite{sitzmann2018saliency}.
  It has been further found in \cite{sitzmann2018saliency}
  that salient objects attract more attention when their number is less or their locations are close. 
  In summary, the content of 360{\dg} video/image has a great impact on attention of subjects.

  \item \textbf{Relationship between HM and EM.} 
  For 360{\dg} video/image, HM reflects the position of the subject's viewport, while EM indicates where the subject fixates on. Thus, the distribution of HM and EM may not be identical.
  Rai \textit{et al.} \cite{rai2017saliency} found an volcano-like distribution of EM within the viewport on 360{\dg} image, showing the difference between the distribution of HM (center of the viewport) and EM. They also statistically evaluated the difference between saliency maps generated from HM and EM for 360{\dg} image. The quantitative results show that the distribution of HM is approximate to but still different from that of EM. As a result, several works 
  developed one model \cite{lebreton2018gbvs360,zhu2018prediction,ling2018saliency} to predict both HM and EM saliency maps, but more works focused on either HM prediction \cite{de2017look,battisti2018feature,nguyen2018your,lebreton2018v,xu2018predicting} or EM prediction \cite{startsev2018360,luz2017saliency,fang2018novel,assens2018scanpath,monroy2018salnet360,chao2018salgan360,suzuki2018saliency,zhang2018saliency}, or they proposed different models to predict HM and EM saliency maps, respectively.
  More details are discussed in the following. 
 
\end{enumerate}

\subsection{Attention modelling approaches}\label{sec:sal:approach}
\begin{figure*}
  \centering
  \includegraphics[width=\textwidth]{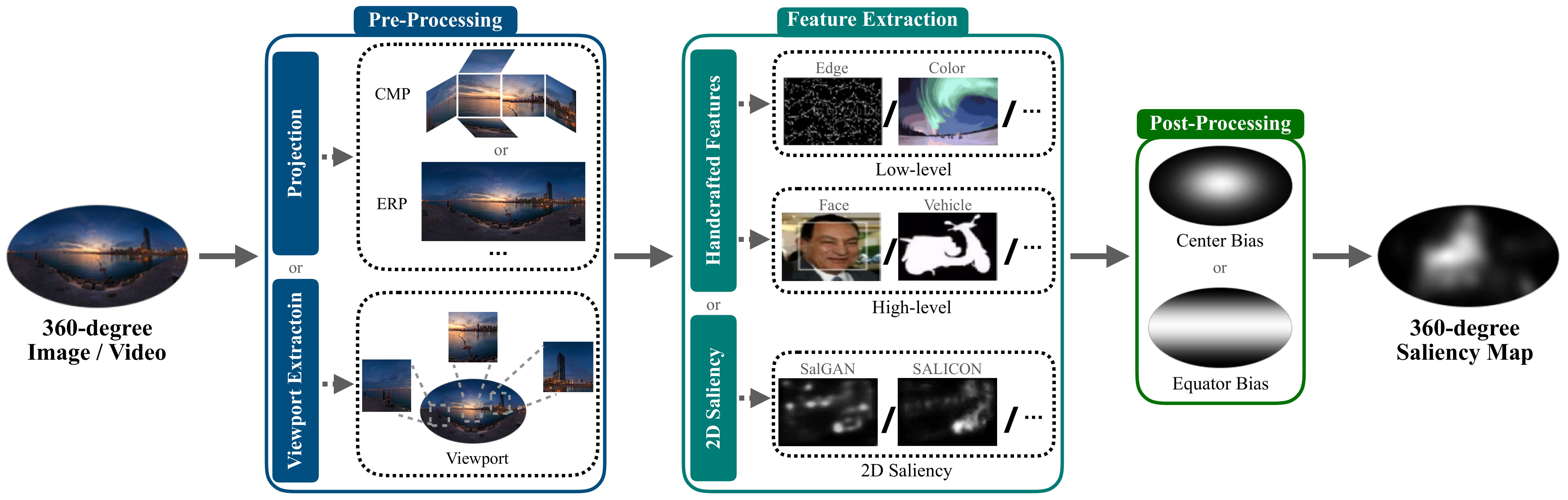}
  \caption{General framework of heuristic saliency approaches on 360{\dg} video/image.}\label{fig:hueristic}
\end{figure*}%
\begin{figure}[t]%
\centering%
\resizebox{\linewidth}{!}{%
\begin{tabular}{|m{7em}|c|}
\hline
  \multicolumn{1}{|c|}{Projection} & Illustration \\
\hline
  Equirectangular Projection (ERP) & \begin{minipage}{0.6\linewidth}
  \includegraphics[width=\linewidth]{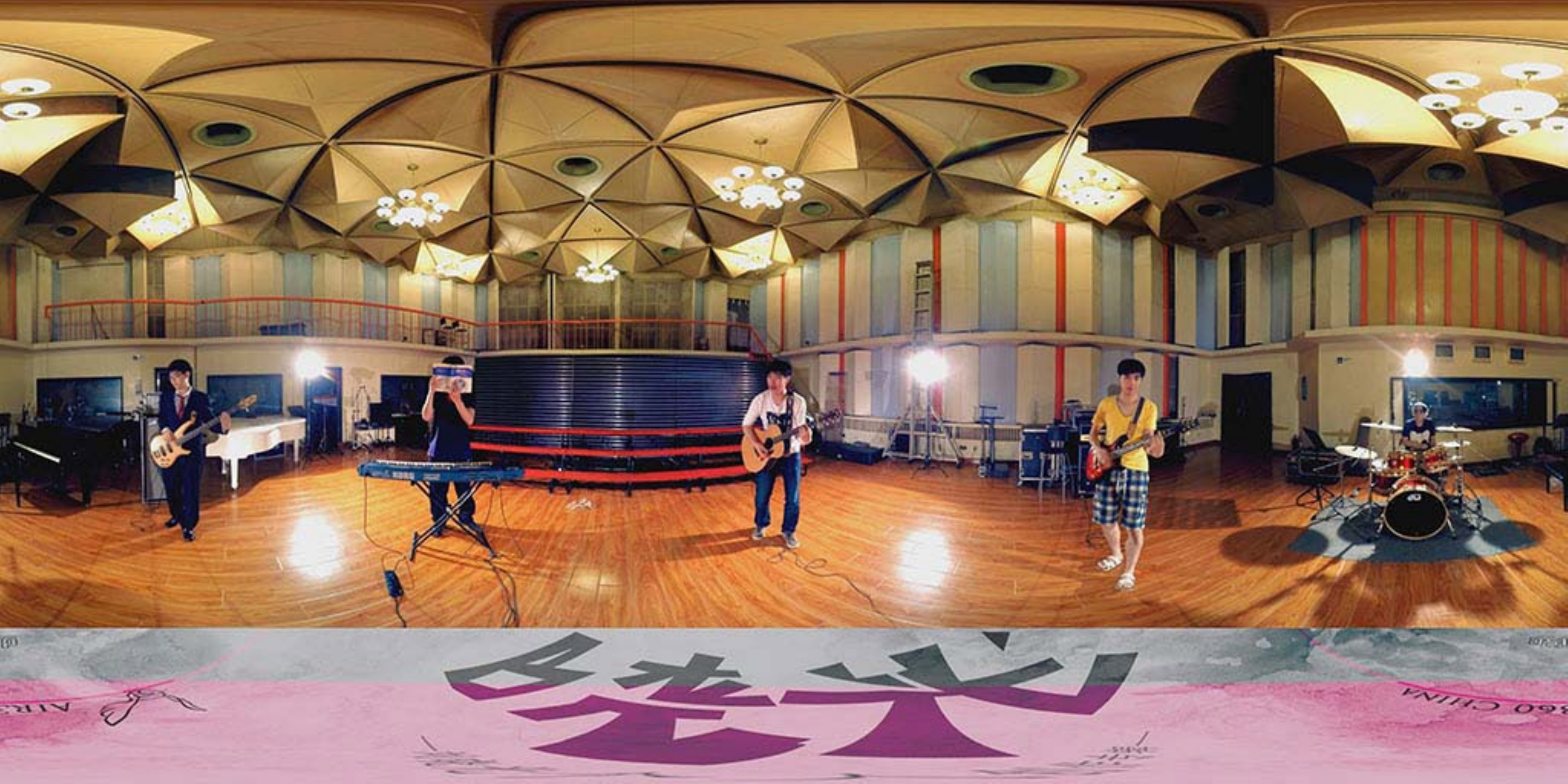}
 \end{minipage}
 \\
  \hline
  Cubemap Projection (CMP) & \begin{minipage}{0.6\linewidth}
  \includegraphics[width=\linewidth]{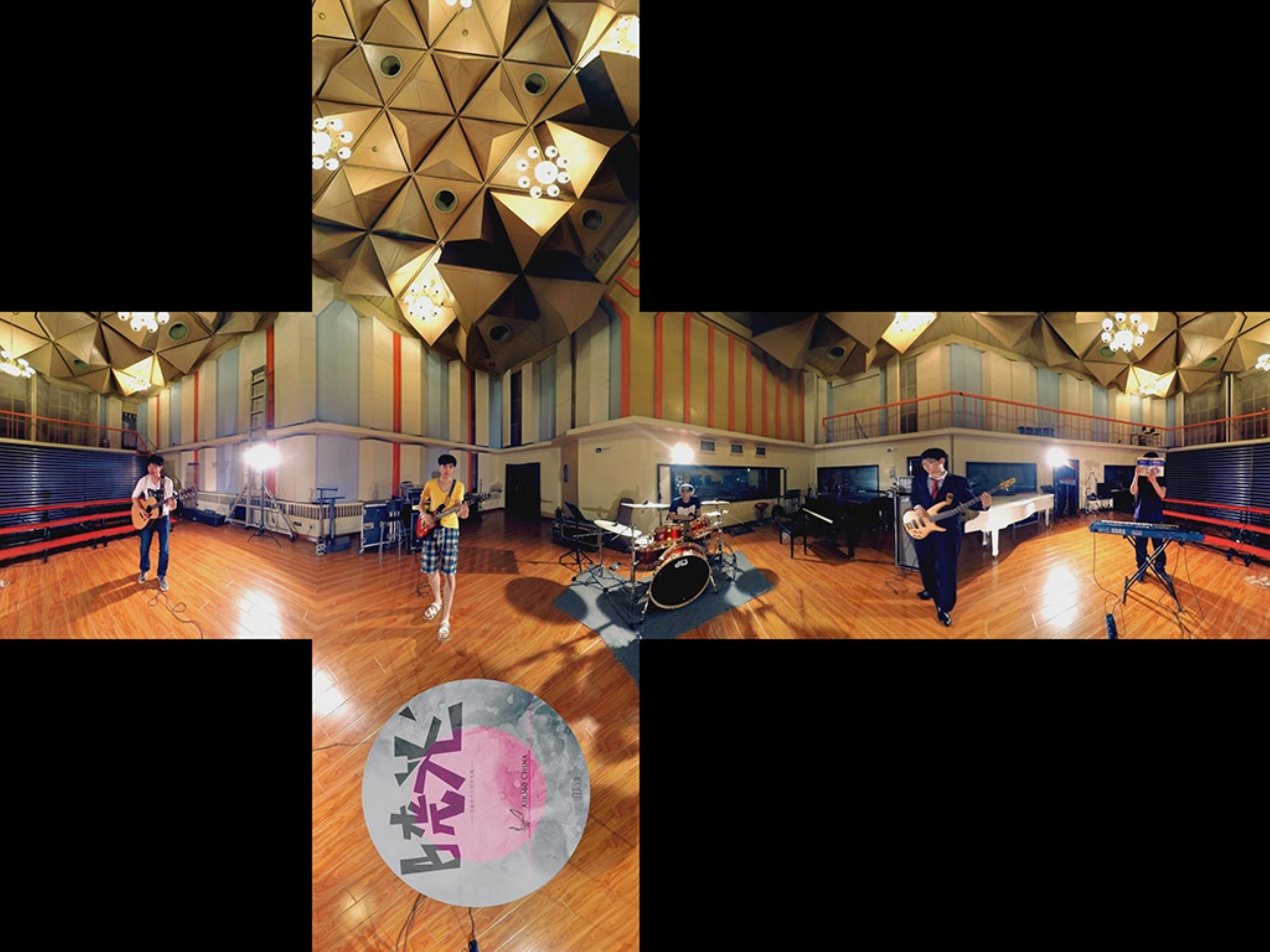}
 \end{minipage}
 \\
  \hline
  Truncated Square Pyramid Projection (TSP) & \begin{minipage}{0.6\linewidth}
  \includegraphics[width=\linewidth]{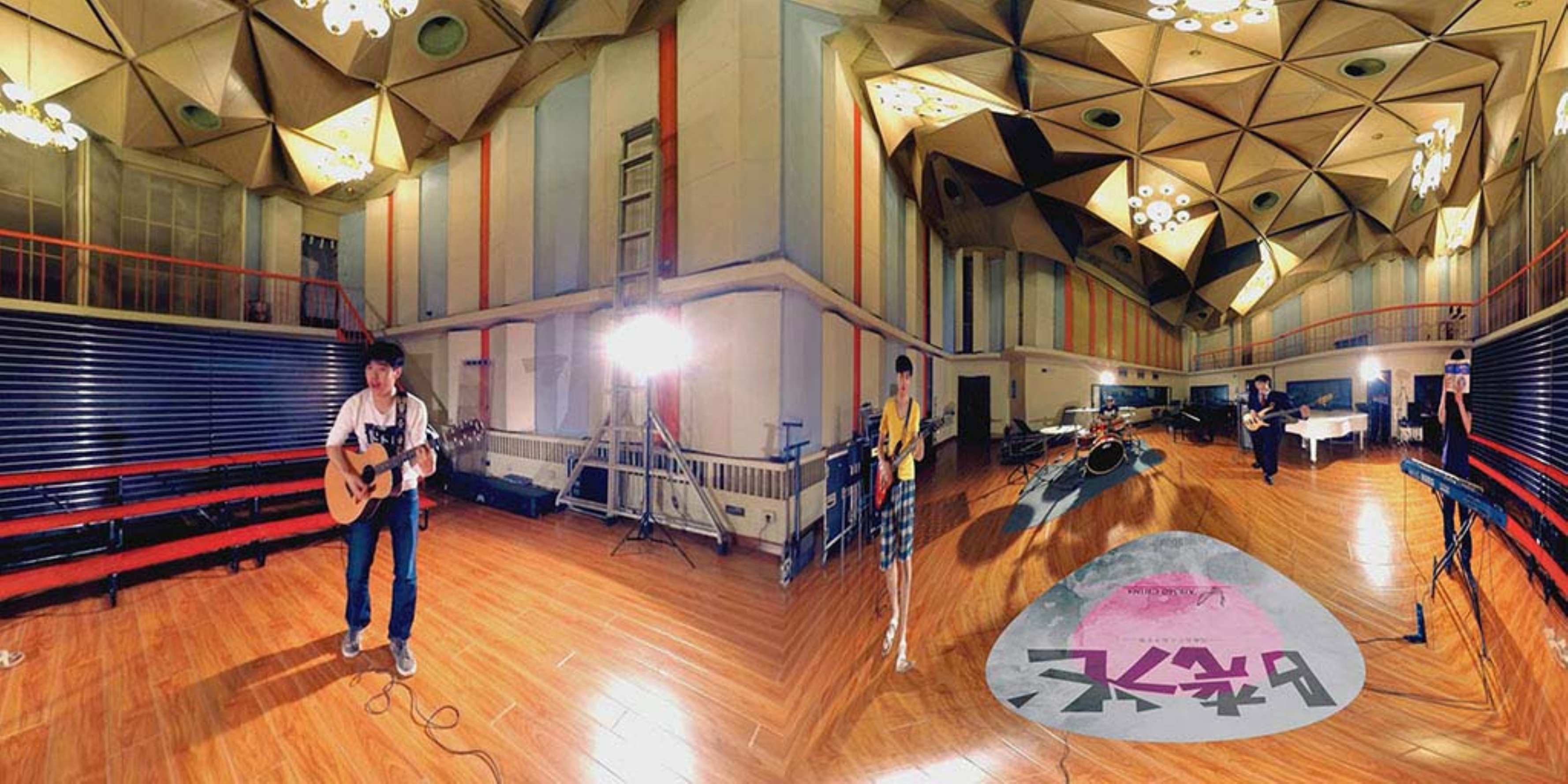}
 \end{minipage}
 \\
 \hline
 Craster Parabolic Projection (CPP) & \begin{minipage}{0.6\linewidth}
  \includegraphics[width=\linewidth]{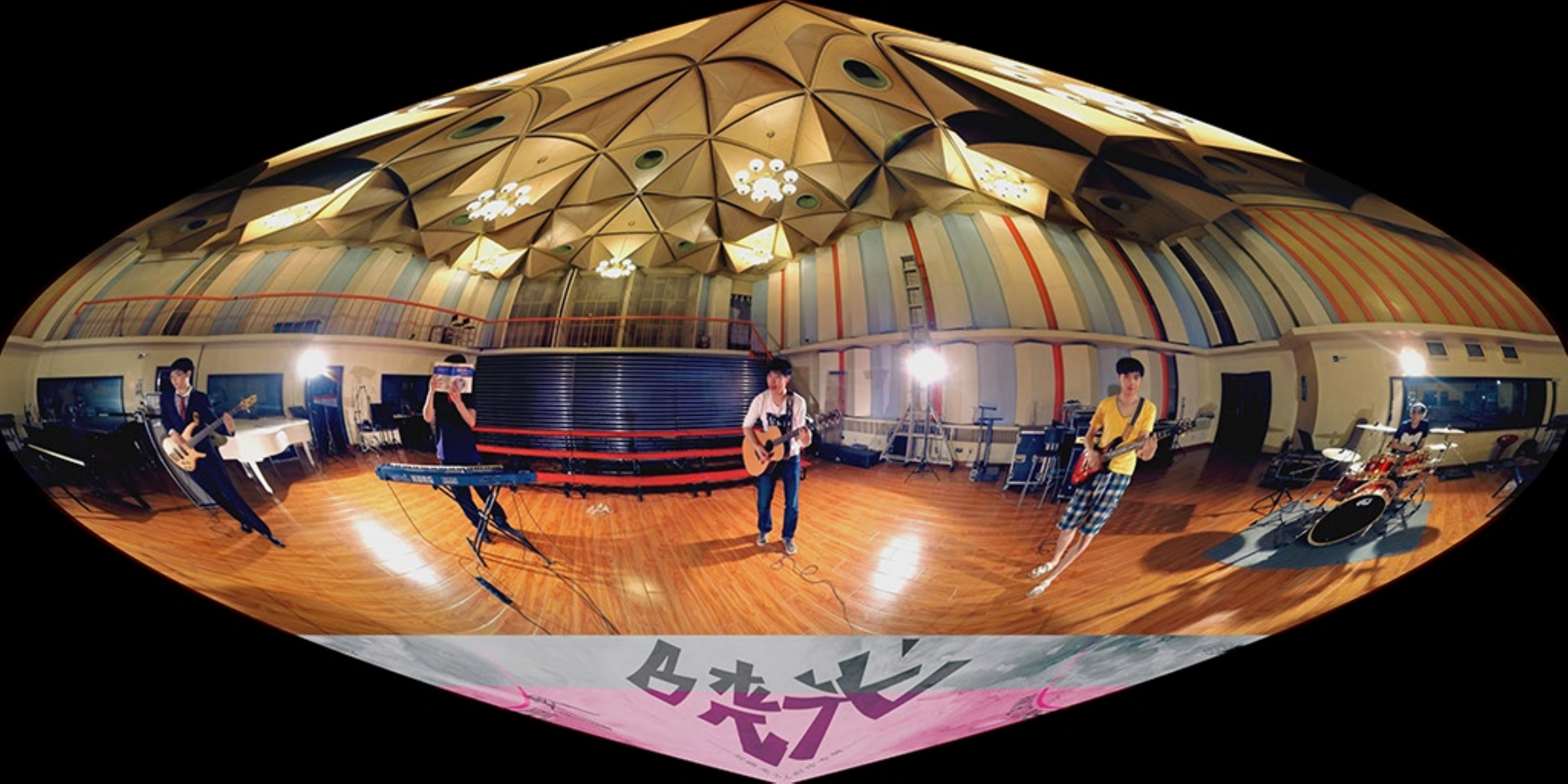}
 \end{minipage}
 \\
  \hline
\end{tabular}
}%
\caption{Examples of several types of projection for 360{\dg} video/image.}\label{fig:sal:proj}%
\end{figure}%
In this section, we review the attention modelling works for 360{\dg} video/image. 
They mainly refer to predicting the HM/EM saliency maps of 360{\dg} video/image, which model the distribution of viewports/RoIs of multiple subjects. Some other attention modelling works are not discussed in this paper, e.g., scanpath prediction \cite{aladagli2017predicting,assens2017saltinet} or HM/EM prediction of an individual subject \cite{fan2017fixation,xu2018gaze,nguyen2018your,xu2018predicting}.
The saliency prediction works can be roughly categorized into heuristic approaches and data-driven approaches. 

\subsubsection{\textbf{Heuristic approaches}}

The heuristic approaches utilize the handcraft features for modelling attention on 360{\dg} video/image. Figure \ref{fig:hueristic} shows an general framework of saliency approaches on 360{\dg} video/image.
The heuristic approaches can be traced back to the year of 2008.
In 2008, Bogdanova \textit{et al.} \cite{bogdanova2008spherical,bogdanova2008visual} proposed
predicting saliency maps on 360{\dg} image through three stages: sphere processing, feature extraction and saliency map generation.
First, \cite{bogdanova2008visual} developed several data processing methods on the sphere, including filtering, Gaussian and Gabor pyramid, up-sampling and normalization on the sphere. These processing methods constituted the foundation of the spherical visual attention models, and even inspired the recent study on the spherical convolutional network \cite{cohen2018spherical}.
Second, \cite{bogdanova2008visual} followed the multi-scale center-surround mechanism \cite{koch1987shifts} of 2D image to extract the attention-related features, and adapted these features to the sphere of 360{\dg} image. 
That is, 7 features of 3 types were extracted: 1 intensity feature, 2 chromatic features and 4 local orientation features. 
Corresponding to the extracted features, 7 spherical conspicuity maps are obtained, and then 3 cue conspicuity maps of intensity, chroma and orientation were calculated from the spherical conspicuity maps. 
Finally, the saliency map was generated by fusing 3 cue conspicuity maps together after normalization.
Afterwards, Bogdanova \textit{et al.} proposed a saliency prediction approach \cite{bogdanova2010dynamic} for 360{\dg} video. In addition to the static saliency map \cite{bogdanova2008visual}, they further proposed calculating the motion saliency maps for 360{\dg} video. To be specific, \cite{bogdanova2008visual} first proposed block matching and motion pyramid calculation methods on the sphere, and then yielded the spherical motion conspicuity maps in the motion magnitude and phase. These conspicuity maps are fused to obtain the motion saliency map of each 360{\dg} video frame. Finally, the dynamic saliency maps are obtained by fusing the static and motion saliency maps. Unfortunately, there had no practical HMD being released at that time, and thus it was impossible to collect HM and EM data of subjects. As a result, quantitative evaluation was not included in \cite{bogdanova2008spherical,bogdanova2008visual,bogdanova2010dynamic}.

Recently, HM and EM data can be easily collected, and thus there have emerged many saliency prediction approaches for 360{\dg} video/image. 
Since saliency prediction approaches have been highly developed for 2D image and video \cite{borji2013state}, it is intuitive to adapt 2D saliency approaches for 360{\dg} video/image via making some modification. However, the adaption of 2D saliency approaches in 360{\dg} video/image may suffer from geometric distortion and border artifacts, caused by sphere-to-plane projections. Figure \ref{fig:sal:proj} illustrates several common projections.
Therefore, several 360{\dg} video/image saliency prediction works \cite{de2017look,lebreton2018gbvs360,maugey2017saliency, startsev2018360} strove to solve these issues.

Abreu \textit{et al.} \cite{de2017look} proposed a fused saliency map (FSM) approach for HM saliency prediction on 360{\dg} image, in which they directly applied SALICON \cite{huang2015salicon} (a 2D image saliency prediction approach) on 360{\dg} image under the sphere-to-plane projection of equirectangular projection (ERP). 
Specifically, the original 360{\dg} image is rotated several different angles and then projected into several 2D images after the ERP mapping, as the variants of the original image. As such, the border artifacts in the two vertical sides of the original image are reduced in the variants.
Then, SALICON is conducted to obtain the saliency maps of all 2D images, which are fused together to generate the final saliency map of the 360{\dg} image.
On the basis of the FSM approach, Lebreton \textit{et al.} \cite{lebreton2018gbvs360} adapted the Boolean map based saliency (BMS) \cite{zhang2016exploiting} approach for 360{\dg} image as the BMS360 approach. A cosine weight is further applied, considering the stretching effect near the horizontal borders of the original 360{\dg} image.

Maugey \textit{et al.} \cite{maugey2017saliency} proposed applying the 2D saliency prediction approach on the faces of 360{\dg} image under the sphere-to-plane projection of cubemap projection (CMP).
CMP has less geometric distortion than ERP, leading to more accurate saliency prediction in 360{\dg} image.
In \cite{maugey2017saliency}, the double cube projection is utilized for mapping the spherical image onto two cubes, with the face center of one cube points towards the corners of the other cube to minimize the overlap between the two CMP-mapped images.
Then, the saliency maps of the faces in the two CMP-mapped images can be obtained by applying traditional 2D saliency prediction approaches.  
The saliency maps are weighted with the following weights:
\begin{equation}\label{eq:sal:face_weight}
  w_\text{face}\left(i_\text{face},j_\text{face}\right)=\frac{1}{1+\left(\frac{\max\left(i_\text{face},j_\text{face}\right)^2}{0.3L_\text{face}}\right)^{10}}\mbox{,}
\end{equation}
to reduce border artifacts.
In the above equation, $(i_\text{face},j_\text{face})$ is the pixel coordinate representing the origin at the center of the cube face; $L_\text{face}$ is the width of the cube face. The final saliency map can be generated by merging the weighted saliency maps.

In addition to \cite{de2017look,lebreton2018gbvs360,maugey2017saliency} that used only one type of projection, Startsev \textit{et al.} \cite{startsev2018360} proposed combining the projections of ERP and CMP for better reducing the negative impact of border artifacts on EM saliency prediction of 360{\dg} image. In order to reduce the artifacts in the two vertical sides of an ERP-mapped 360{\dg} image, a variant of the 360{\dg} image is generated by swapping the left and right halves of the image. Then, 2D saliency prediction approaches are applied on both the origin image and its variant, such that 2 saliency maps can be obtained.
In order to reduce the artifacts in the two horizontal sides, CMP is incorporated in \cite{startsev2018360}. Specifically, after projecting the original 360{\dg} image through CMP,  2D saliency prediction approaches are applied to obtain 2 saliency maps, corresponding to the top and bottom faces of the cube. The two cube faces are padded with adjacent faces to reduce the border artifacts in the faces. Finally, 2 saliency maps from ERP and 2 saliency maps from CMP are fused together via pixel-wise maximum in ERP. The equator bias mentioned in Section \ref{sec:sal:ds} is also incorporated in \cite{startsev2018360} as a post-processing step.

Given a spherical location, the viewport image corresponding to this location can be obtained with the sphere-to-plane gnomonic projection \cite{snyder1987map}. 
For 360{\dg} video/image, some other works \cite{lebreton2018gbvs360, luz2017saliency} utilized 2D saliency prediction approaches on the viewport images, rather than the projected 2D images. 
The advantage is that the extracted viewport images of 360{\dg} video/image are of little geometric distortion.
However, the difficulty is how to integrate the saliency maps of different viewports together for generating the final spherical saliency maps.
Lebreton \textit{et al.} \cite{lebreton2018gbvs360} utilized the graph-based visual saliency (GBVS) \cite{harel2006graph} approach in the proposed GBVS360 approach. GBVS is applied on several extracted viewport images to avoid the geometric distortion of ERP when extracting Gabor-based features. Subsequently, the saliency maps of viewports are projected back to the ERP plane and then fused together for obtaining the final saliency mpas of both HM and EM. 
Luz \textit{et al.} proposed an saliency detection model (SDM) approach \cite{luz2017saliency} for modelling EM on 360{\dg} image, in which the viewport images are extracted at spherical locations corresponding to all pixels. Then, the authors applied the pre-trained multi-level network (ML-Net) \cite{cornia2016deep} on viewport images, obtaining viewport saliency maps. The viewport saliency maps are integrated together weighted by the equator bias, yielding the final EM saliency map of 360{\dg} image.

Apart from applying 2D saliency prediction approaches on 360{\dg} video/image, some works \cite{battisti2018feature,zhu2018prediction, fang2018novel} have been proposed to extract handcraft features for saliency prediction of 360{\dg} video/image.
Specifically, Battisti \textit{et al.} \cite{battisti2018feature} proposed an approach (named RM3) for HM saliency prediction on 360{\dg} image. Both low- and high-level feature maps are obtained based on the viewport images. 
In \cite{battisti2018feature}, the low-level maps of viewport images are based on the integration of some low-level features: hue, saturation and GBVS
features.
For high-level features, skin \cite{thakur2011face} and face \cite{viola2004robust} are detected in each viewport for generating the high-level feature maps. Finally, the low- and high-level maps are integrated to obtain the final saliency map.
Zhu \textit{et al.} \cite{zhu2018prediction} proposed an approach for 360{\dg} image HM and EM saliency prediction, in which low- and high-level feature maps are also extracted from viewport images. Low-level features refer to orientations, frequency bandwidths, color channels and symmetrical balance, while the high-level features include cars and people detected in the viewport images. Similar to \cite{battisti2018feature}, \cite{zhu2018prediction} produces the final saliency map of 360{\dg} image through the integration of both low- and high-level features. 
Different from \cite{battisti2018feature,zhu2018prediction}, Fang \textit{et al.} \cite{fang2018novel} proposed extracting the handcraft low-level features of the whole 360{\dg} image for EM saliency prediction. Specifically, the input 360{\dg} image is first segmented into super-pixels \cite{achanta2012slic} at different levels. Then, the low-level features of luminance, texture, color and boundary connectivity are extracted on the super-pixels, which are used to generated the final saliency maps.

\subsubsection{\textbf{Data-driven approaches}} 
The visual attention models for 360{\dg} video/image can be also learned from the training data, benefiting from the existing datasets.
Along with the established datasets, several saliency prediction approaches have been proposed for predicting saliency maps of either 360{\dg} image \cite{ling2018saliency, assens2018scanpath, monroy2018salnet360, chao2018salgan360, nguyen2018your} or 360{\dg} video \cite{suzuki2018saliency, lebreton2018v, cheng2018cube, zhang2018saliency, xu2018predicting}.

For 360{\dg} image, Ling \textit{et al.} proposed the color dictionary based sparse representation (CDSR) approach \cite{ling2018saliency} for HM and EM saliency prediction. First, an over-complete color dictionary is trained over the 2D images of MIT1003 \cite{judd2009learning}. 
In the test stage, the input 360{\dg} image is first divided into sub-images. Subsequently, the non-overlapping patches are extracted from the sub-images, and the sparse representations of the extracted patches are obtained with the trained color dictionary.
For each patch, the center-surround differences between itself and other patches are calculated via the $\ell_2$ norm of the differences between their sparse representations.
The patches are further weighted by human visual acuity to obtain their saliency maps. The saliency maps of all patches are finally fused together, and then combined with an equator bias map to obtain the overall saliency map of the input 360{\dg} image.

In the new era of deep learning \cite{goodfellow2016deep}, there have emerged many effective architectures of deep neural networks (DNNs), such as convolutional neural network (CNN), generative adversarial network (GAN), and long short-term memory (LSTM) network. The DNN-based approaches also achieve great success in saliency prediction for 2D image and video \cite{borji2018saliency}. Such great success can also be brought to advance saliency prediction for 360{\dg} video/image. Assens \textit{et al.} proposed a CNN-based approach (called SaltiNet) \cite{assens2018scanpath} with the same architecture as a CNN-based 2D image saliency prediction approach (called SalNet) \cite{pan2016shallow}, except the last layer for EM saliency prediction on 360{\dg} image.
The SaltiNet model is initialized with the pre-trained parameters of SalNet \cite{pan2016shallow} and then trained over the Salient360 dataset.
For training the SaltiNet model, the binary cross entropy (BCE) loss is applied, which is the same as that in 2D saliency prediction \cite{pan2017salgan}.

Monroy \textit{et al.} proposed a SalNet360 \cite{monroy2018salnet360} approach, applying CNN on the cube faces of 360{\dg} image under CMP. In \cite{monroy2018salnet360}, the first stage has the same architecture as SalNet, extracting the 2D saliency maps of the cube faces. Then, the spherical coordinates of the cube faces are concatenated with the extracted saliency maps, flowing into the second stage of SalNet360 with a fully convolutional network (FCN) architecture. The outputs of SalNet360 are converted back to the ERP-projected plane to be fused together for obtaining the final saliency map. The first stage of SalNet360 is trained on SALICON \cite{huang2015salicon}, while the second stage is trained on Salient360 with random selected viewports, both with the Euclidean loss function.

In addition to CNN, Chao \textit{et al.} \cite{chao2018salgan360} proposed applying GAN on cube faces. Different from \cite{monroy2018salnet360}, Chao \textit{et al.} \cite{chao2018salgan360} proposed generating several 2D images from an input 360{\dg} image by rotating the cubes at different CMPs. In \cite{chao2018salgan360}, the Saliency GAN (SalGAN) \cite{pan2017salgan} model is fine-tuned over Salient360, such that \cite{chao2018salgan360} is named as SalGAN360. Since Kullback–Leibler divergence (KLD), PLCC and normalized scanpath saliency (NSS) are widespreadly applied in the evaluation of saliency prediction, These metrics are combined together in the loss function of the generator in SalGAN360. Finally, the saliency maps of 360{\dg} images under different CMPs are all converted back to the ERP plane and then fused together for producing the final saliency map of the input 360{\dg} image.

Suzuki \textit{et al.} \cite{suzuki2018saliency} proposed applying CNN on the extracted viewports of 360{\dg} image, rather than the cube faces of \cite{monroy2018salnet360, chao2018salgan360}.
Different from the previous works, \cite{suzuki2018saliency} proposed to learn the equator bias with a layer in its CNN architecture.
Finally, the saliency maps of viewports are fused together and masked with the learned equator bias, for generating the final saliency map of 360{\dg} image.

For 360{\dg} video, Nguyen \textit{et al.} \cite{nguyen2018your} applied a 2D image saliency prediction approach (\textit{i.e.}, SalNet) in their PanoSalNet approach \cite{nguyen2018your} to predict HM saliency map of each frame. 
However, the PanoSalNet model does not take into account the temporal features in saliency prediction.
The model of PanoSalNet is fine-tuned over 360{\dg} video datasets of Corbillon \textit{et al.} \cite{corbillon2017360} and Wu \textit{et al.} \cite{wu2017dataset},
such that the PanoSalNet approach can be used for 360{\dg} video saliency prediction.
In contrast to \cite{nguyen2018your}, other advanced approaches \cite{lebreton2018v, cheng2018cube, zhang2018saliency, xu2018predicting} consider the temporal features of 360{\dg} video in HM/EM saliency prediction.

Specifically, Lebreton \textit{et al.} proposed extending BMS360 to V-BMS360 \cite{lebreton2018v}, which extracts two motion-based features for 360{\dg} video, both based on optical flow.
One motion-based feature is motion source, while the other one is called motion surroundness.
The motion source feature is extracted with random walkers \cite{revesz2005random} moving in the opposite direction of the MV at the place where they are located.
For extracting the motion surroundness feature, the optical flow image is normalized to reduce the impact of non-uniform sampling alongside latitude in ERP. Then, BMS360 is applied to the normalized optical flow image for extracting the motion surroundness feature. 
For adaptively fusing the two motion-based features, a CNN is trained to classify the camera motion type of the video.
In post-processing, the authors proposed cutting off the saliency values located near the two vertical sides of the 360{\dg} video, which is similar to the center bias of 360{\dg} video discussed in Section \ref{sec:sal:ds}.

Cheng \textit{et al.} \cite{cheng2018cube} proposed a DNN-based saliency prediction approach for 360{\dg} video, consisting of both static and temporal models. 
The static model predicts the saliency map for each single frame, while the temporal model adjusts the outputs of the static model based on temporal features.
The static model predicts the saliency maps on the cube faces of the input 360{\dg} video under CMP. However, the cube padding, instead of the zero padding, is applied, despite the fact that zero padding is generally adopted in CNN. In the cube padding, the cube faces are padded with the content of adjacent cube faces, in order to reduce the border artifacts and preserve the receptive field of neurons across 360{\dg} content. 
The temporal model utilizes the network of convolutional LSTM (ConvLSTM), since 360{\dg} video can be regarded as a sequence of 360{\dg} frames. The DNN model of \cite{cheng2018cube} is trained with 3 proposed temporal loss functions, considering the invariance, smoothness and motion in spatial and temporal neighbourhood.

\begin{table*}[t]
  \centering
  \caption{Summary of EM/HM saliency prediction approaches for 360{\dg} video/image.}\label{tab:sal:approach}%
\resizebox{\textwidth}{!}{%
\begin{tabular}{|c|c|c|c|c|c|c|c|}
\hline
\multicolumn{2}{|c|}{Approach} & Image/Video & HM/EM & Bias & Metrics & Benchmark & Availability \\
\hline
\multirow{10}{*}{Heuristic} & \multirow{2}{*}{Bogdanova \textit{et al.}} & Image \cite{bogdanova2008spherical,bogdanova2008visual} & --- & --- & --- & --- & No \\
\cline{3-8}  &   & Video \cite{bogdanova2010dynamic} & --- & --- & --- & --- & No \\
\cline{2-8}  & FSM \cite{de2017look} & Image & HM & --- & ROC, AUC-Borji & Abreu \textit{et al.} \cite{de2017look} & No \\
\cline{2-8}  & GBVS360, BMS360, ProSal \cite{lebreton2018gbvs360} & Image & HM \& EM & Equator & KLD, PLCC, NSS, AUC-Judd & Salient360 \cite{gutierrez2018toolbox} & Yes \\
\cline{2-8}  & Maugey \textit{et al.} \cite{maugey2017saliency} & Image & --- & --- & --- & --- & Yes \\
\cline{2-8}  & Startsev \textit{et al.} \cite{startsev2018360} & Image & EM & Equator & KLD, PLCC, NSS, AUC & Salient360 \cite{gutierrez2018toolbox} & Yes \\
\cline{2-8}  & SDM \cite{luz2017saliency} & Image & EM & Equator & RMSE & Salient360 \cite{gutierrez2018toolbox} & No \\
\cline{2-8}  & RM3 \cite{battisti2018feature} & Image & HM & Equator & KLD, PLCC & Salient360 \cite{gutierrez2018toolbox} & No \\
\cline{2-8}  & Zhu \textit{et al.} \cite{zhu2018prediction} & Image & HM \& EM & Equator & KLD, PLCC, NSS, AUC-Judd & Salient360 \cite{gutierrez2018toolbox} & No \\
\cline{2-8}  & Fang \textit{et al.} \cite{fang2018novel} & Image & EM & Equator & KLD, PLCC & Salient360 \cite{gutierrez2018toolbox} & No \\
\hline
\multirow{10}{*}{Data-driven} & CDSR \cite{ling2018saliency} & Image & HM \& EM & Equator & KLD, PLCC, NSS, ROC & Salient360 \cite{gutierrez2018toolbox} & No \\
\cline{2-8}  & SaltiNet \cite{assens2018scanpath} & Image & EM & --- & KLD, PLCC, NSS, ROC & Salient360 \cite{gutierrez2018toolbox} & Yes \\
\cline{2-8}  & SalNet360 \cite{monroy2018salnet360} & Image & EM & --- & KLD, PLCC, NSS, AUC & Salient360 \cite{gutierrez2018toolbox} & Yes \\
\cline{2-8}  & SalGAN360 \cite{chao2018salgan360} & Image & EM & --- & KLD, PLCC, NSS, AUC & Salient360 \cite{gutierrez2018toolbox} & Yes \\
\cline{2-8}  & Suzuki \textit{et al.} \cite{suzuki2018saliency} & Image & EM & Equator & KLD, PLCC, NSS, AUC & Salient360 \cite{gutierrez2018toolbox} & No \\
\cline{2-8}  & PanoSalNet \cite{nguyen2018your} & Video & HM & --- & PLCC, NSS, sAUC & Corbillon \textit{et al.} \cite{corbillon2017360} and Wu \textit{et al.} \cite{wu2017dataset} & Yes \\
\cline{2-8}  & V-BMS360 \cite{lebreton2018v} & Video & HM & Center & KLD, PLCC, NSS, AUC-Judd & Salient360 \cite{david2018dataset} & Yes \\
\cline{2-8}  & Cheng \textit{et al.} \cite{cheng2018cube} & Video & --- & --- & PLCC, AUC-Judd, AUC-Borji & Wild-360 \cite{cheng2018cube} & Yes \\
\cline{2-8}  & Spherical U-Net \cite{zhang2018saliency} & Video & EM & --- & PLCC, NSS, AUC-judd & Zhang \textit{et al.} \cite{zhang2018saliency} & Yes \\
\cline{2-8}  & DHP \cite{xu2018predicting} & Video & HM & Center & PLCC, NSS, sAUC & PVS-HM \cite{xu2018predicting} & Yes \\
\hline
\end{tabular}%
}%
\end{table*}%
The aforementioned approaches directly apply planar CNN on saliency prediction for 360{\dg} video/image.
In contrast, Zhang \textit{et al.} \cite{zhang2018saliency} proposed a new type of spherical CNN for 360{\dg} content. In the spherical CNN of \cite{zhang2018saliency}, the kernel is defined on a spherical crown, and the convolution corresponds to the rotation of the kernel on sphere. The spherical CNN \cite{zhang2018saliency} is implemented under the plane of ERP, and the kernel is re-sampled to this plane. Based on the spherical CNN, a spherical U-Net was proposed for EM saliency prediction on 360{\dg} video, and its inputs include one frame and the predicted saliency maps of several previous frames for better modelling dynamic saliency. Based on traditional mean squared error (MSE) loss, the authors also defined a spherical MSE (S-MSE) loss function for training the spherical U-Net to reduce the non-uniform sampling of ERP. Assume that one predicted saliency map is $\hat{S}$ and its ground-truth is $S$, which are defined on the unit sphere with longitude and latitude of $(\theta,\phi)$. Then, S-MSE can be calculated by
\begin{equation}\label{eq:sal:smse}
  \mathcal{L}_\text{S-MSE}=\sum_{\theta=0,\phi=0}^{\Theta,\Phi}\frac{\Omega(\theta,\phi)}{4\pi}\left(S(\theta,\phi)-\hat{S}(\theta,\phi)\right)^2\mbox{,}
\end{equation}
where $\Omega(\theta,\phi)$ denotes the solid angle of the sampled area on the corresponding saliency map located at $(\theta,\phi)$ under ERP.

Xu \textit{et al.} \cite{xu2018predicting} proposed an HM saliency prediction approach for 360{\dg} video, which utilizes deep reinforcement learning (DRL), instead of supervised learning in the traditional works. With the proposed DRL-based HM prediction (DHP) approaches, an \textit{agent} is trained with the accumulated \textit{reward} on its \textit{actions} of HM on 360{\dg} video, so that the \textit{agent} can mimic the long-term HM behavior of one subject.
Note that the \textit{reward} in the training stage is designed to measure the difference between the predicted \textit{actions} of the \textit{agent} and the actual HM of the subject.
Then, a new framework of multi-workflow DRL was developed, in which each DRL workflow predicts the HM data of one subject, and the multiple workflows yield the HM data of several subjects. Consequently, the saliency maps of 360{\dg} video can be obtained by convoluting the HM data of all DRL workflows.

The evaluation metrics for EM/HM saliency prediction on 360{\dg} video/image are the same as those of 2D saliency prediction, including PLCC, NSS, KLD, area under ROC curve (AUC) etc. Thanks to the Salient360 grand challenges\footnote{\url{https://salient360.ls2n.fr/}}, there are several works evaluating their approaches on the Salient360 dataset, making these approaches comparable. Others evaluated their approaches on their own datasets. Table \ref{tab:sal:approach} summarizes the above reviewed approaches for saliency prediction on 360{\dg} video/image.

\section{Visual Quality Assessment}
There are two types of VQA, \textit{i.e.}, subjective and objective VQA. In subjective VQA, subjects are required to rate scores for the viewed images or videos, so that their subjective quality scores can be obtained. Since human is the ultimate receiver of video/image, the subjective quality scores can be seen as the ground truth quantitative representation of the visual quality. For objective VQA, many approaches have been proposed to model the subjective quality scores of video/image. As mentioned in Section \ref{sec:sal}, the viewing mechanism is different between 2D and 360{\dg} video/image, leading to different experience. Therefore, the subjective VQA experiment methods and objective VQA approaches for 2D video/image are not appropriate for 360{\dg} video/image. In this section, we survey on the works about VQA on 360{\dg} video/image, from the aspects of subjective and objective VQA. 
The subjective VQA can be directly used to obtain the subjective scores, and thus it is normally used to establish the dataset for training and evaluating the objective VQA approaches. In the following, we introduce subjective VQA methods and datasets together.
\subsection{Subjective VQA and datasets}\label{sec:vqa:ds}
Considering the unique viewing mechanism of 360{\dg} video/image, there are several works designing dedicated subjective VQA methods, mostly modified from those for 2D content \cite{itur2012methodology}. We survey them from the following aspects.
\subsubsection{\textbf{Test material display}}
For 2D video/image, the test materials are displayed by flat monitors. However, 360{\dg} video/image should not be directly displayed by flat monitors under projection, because of the geometric distortion (\textit{e.g.}, ERP) and spatial disorganization (\textit{e.g.}, CMP). To display 360{\dg} video/image with flat monitors, Zakharchenko \textit{et al.} \cite{zakharchenko2016quality} proposed rendering random viewports for 360{\dg} video/image as the test materials, and different subjects view the same materials in the experiments. In contrast, Boyce \textit{et al.} \cite{boyce2017jvet} and Hanhart \textit{et al.} \cite{hanhart2018360} proposed to define certain static viewports for different projection types, and these viewports are rendered as the test materials. This is mainly for checking discontinuous edges caused by projection.

Since 360{\dg} video/image is mostly viewed in the HMD, it is intuitive to display test materials with the HMD in subjective experiments for providing the identical environment as far as possible. However, the difficulty is how to record the data without making subjects take off the HMD during the experiment. In some works \cite{huang2018modeling,duan2017ivqad,zhang2017subjective,lopes2018subjective,singla2017comparison}, the subjective data were recorded manually by an assistant during the experiment. In contrast, Upenik \textit{et al.} \cite{upenik2016testbed} proposed a testbed for subjective experiment of 360{\dg} video/image using the HMD, in which a software was developed with the functions of 360{\dg} video/image playback, subjective score rating, experiment instruction and HM data collection. Similarly, there are many other works \cite{sun2017cviqd,sun2018large,duan2018perceptual,xu2018assessing,li2018bridge} that developed various software solving the problems of data collection and process control. Different from the subjective experiments with flat monitors, there is little restrict on the conditions of experiment with the HMD, since the environment factors, such as room illumination and distance to the screen, have no impact on viewing 360{\dg} video/image with the HMD. However, free viewing needs to be allowed for subjects, and thus it is proposed in these works that subjects stand \cite{huang2018modeling} or sit on a swivel chair \cite{upenik2016testbed,upenik2017performance,sun2017cviqd,sun2018large,duan2017ivqad,duan2018perceptual,lopes2018subjective,singla2017comparison,xu2017subjective,xu2018assessing,li2018bridge} during the subjective experiment.
\subsubsection{\textbf{Test method}}
There are 3 key factors defined in the test method: session setting, test material arrangement and rating scale. Since 360{\dg} video/image is a new type of multimedia content, it is universally acknowledged that a training session is indispensable before the test session in the subjective experiment, for making subjects familiar with 360{\dg} content and even the HMD. In \cite{zhang2017subjective}, there is an extra pre-test session between the training and test sessions, in order to guarantee the robustness of the rated quality scores.

For the test material arrangement, the majority of works directly applied those designed for 2D video/image. Specifically, many works \cite{upenik2016testbed,upenik2017performance,sun2017cviqd,sun2018large,huang2018modeling,duan2017ivqad,duan2018perceptual,chen2018towards,zhang2018subjective,lopes2018subjective,xu2017subjective,xu2018assessing,tran2018study,li2018bridge} adopted the absolute category rating (ACR) and ACR with hidden reference (ACR-HR) methods \cite{itut2008subjective}, which are also known as the single-stimulus (SS) methods \cite{itur2012methodology}. In these methods, only one test material is presented at a time and the test materials are presented only once in a random order. As such, subjects can rate for each test material without direct comparison. Singla \textit{et al.} \cite{singla2017comparison} proposed a modified ACR (M-ACR) method, in which each test material is presented twice before rating. In addition, Zhang \textit{et al.} \cite{zhang2017subjective} proposed a method for fine-grained subjective VQA on 360{\dg} video modified from the subjective assessment of multimedia video quality (SAMVIQ) \cite{itur2007methodology} method for 2D video, which is called subjective assessment of multimedia panoramic video quality (SAMPVIQ). In SAMPVIQ, impaired sequences with the same content are grouped in a scene, which are then divided into several non-overlapping groups. The experiment is implemented group-by-group and scene-by-scene. For the assessment on one group, both explicit and hidden reference sequences are included for collecting the quality score of the reference sequence but still providing a baseline for comparison. The subject is allowed to view test sequences any times in any order. After the subject finishing rating scores for all test sequences in the group, the experiment moves on the next group. As a result, the sequences of different quality levels with tiny difference can be distinguished.

For the rating scale, either discrete or continuous quality scale is adopted in the above works, following the definition for 2D video/image \cite{itur2007methodology,itut2008subjective,itur2012methodology}.
\begin{table*}[t]
  \centering
  \caption{Summary of 360{\dg} image or video datasets for VQA.}\label{tab:vqa:ds}%
  \resizebox{\textwidth}{!}{%
    \begin{tabular}{|c|c|c|c|m{5em}|c|m{11em}|m{30em}|m{5.5em}|}
\hline
Dataset & Image/Video & Subjects & Dataset Size * & \multicolumn{1}{c|}{Resolution} & Duration \# & \multicolumn{1}{c|}{Distortion} & \multicolumn{1}{c|}{Description} & \multicolumn{1}{c|}{Availability} \\
\hline
Upenik \textit{et al.} 2016 \cite{upenik2016testbed} & Image & \tabincell{c}{48\\Ages: 19-36} & 6/60 & 3000x1500 (ERP)  and 2250x1500 (CMP) & \tabincell{c}{30 s\\in average} & JPEG with different quality factors (QFs); different projection & HMD: MergeVR + iPhone 6S. The ACR-HR method with 5 quality levels was adopted in the experiment. The subjects were equally divided into 2 groups. One group of subjects viewed sequences under ERP, while the other group of subjects viewed sequences under CMP. MOS values and HM data of subjects are also included. & \multicolumn{1}{c|}{N/A} \\
\hline
Upenik \textit{et al.} 2017 \cite{upenik2017performance} & Image & \tabincell{c}{45\\Ages: 18-32} & 4/104 & 3000$\times$1500 pixels & --- & JPEG and JPEG 2000 with different QFs; H.265 with different bitrate; different projection & The testbed and condition is the same as \cite{upenik2016testbed}. MOS values and HM data of subjects are also included. & \multicolumn{1}{c|}{N/A} \\
\hline
CVIQD \cite{sun2017cviqd} & Image & 20 & 5/170 & 4096$\times$2048 pixels & --- & JPEG with different QFs; H.264 and H.265 with different quantization parameters (QPs) & HMD: HTC Vive. The SS method with 10 quality levels was adopted in the experiment. MOS values are also calculated. & \multicolumn{1}{c|}{N/A} \\
\hline
CVIQD2018 \cite{sun2018large} & Image & \tabincell{c}{20\\Ages: 21-25} & 16/544 & 4096$\times$2048 pixels & --- & JPEG with different QFs; H.264 and H.265 with different QPs & HMD: HTC Vive. The SS method with 10 quality levels was adopted in the experiment. MOS values are also calculated. & \multicolumn{1}{c|}{N/A} \\
\hline
Huang \textit{et al.}  \cite{huang2018modeling} & Image & \tabincell{c}{98\\Ages: 18-25} & 12/156 & 9104$\times$4552 pixels & 20 s & JPEG with different QFs; different resolution & HMD: HTC Vive. The ACR method with continuous quality scale was adopted in the experiment. Each subject viewed 36 images and each sequence was viewed by at least 16 subjects. & \url{http://vision.nju.edu.cn/index.php/data-base/item/64-im-images} \\
\hline
OIQA \cite{duan2018perceptual} & Image & 20 & 16/336 & 11332$\times$5666 to 13320$\times$6660 pixels & 20 s & JPEG and JPEG2000 with different QFs; different levels of Gaussian blur and white Gaussian noise & HMD: HTC Vive. Eye-tracker: aGlass. The SS method with 5 quality levels and explicit references was adopted in the experiment. MOS values, HM and EM data of subjects are also included. & \multicolumn{1}{c|}{N/A} \\
\hline
IVQAD2017 \cite{duan2017ivqad} & Video & 13 & 10/160 & 4096$\times$2048 pixels & 15 s & MPEG-4 with different bitrate; different resolution; different frame rate & HMD: HTC Vive. The SS method with 5 quality levels and explicit references was adopted in the experiment. MOS values are also calculated. & \multicolumn{1}{c|}{N/A} \\
\hline
Y.Zhang \textit{et al.} \cite{zhang2018subjective} & Video & \tabincell{c}{30\\Ages: 20-26} & 10/60 & 3600$\times$1800 pixels & 10 s & H.265 with different QPs & HMD: HTC Vive. The ACR-HR method with 5 quality levels was adopted in the experiment. DMOS values and HM data of subjects are also included. & \multicolumn{1}{c|}{N/A} \\
\hline
B.Zhang \textit{et al.} \cite{zhang2017subjective} & Video & \tabincell{c}{23\\Averate age: 22.3} & 16/400 & 4096$\times$2048 pixels & --- & VP9, H.264 and H.265 with different bitrate; different levels of Gaussian noise and box blur & HMD: HTC Vive. The SAMPVIQ method proposed in this paper was adopted in the experiment. MOS and DMOS values are also calculated. & \multicolumn{1}{c|}{Upon request} \\
\hline
Lopes \textit{et al.} \cite{lopes2018subjective} & Video & \tabincell{c}{37\\Ages: 22-55} & 6/85 & 8192$\times$4096 and 3840$\times$1920 pixels & 10 s & H.265 with different QPs; different resolution and frame rate & HMD: Oculus Rift. The ACR-HR and ACR methods with 5 quality levels were adopted in the experiment. MOS and DMOS values are also calculated. & \multicolumn{1}{c|}{Upon request} \\
\hline
Singla \textit{et al.} \cite{singla2017comparison} & Video & \tabincell{c}{30\\Ages: 19-36} & 6/66 & 1080P and 4K & 10 s & H.265 with different bitrate & HMD: Oculus Rift CV1. The M-ACR method with 5 quality levels proposed in this paper was adopted in the experiment. MOS values and HM data of subjects are also included. & \multicolumn{1}{c|}{N/A} \\
\hline
VR-VQA48 \cite{xu2017subjective,xu2018assessing} & Video & 48 & 12/48 & 4096$\times$2048 pixels & 12 s & H.265 with different QPs & HMD: HTC Vive. The SS method with continuous quality scale was adopted in the experiment. MOS values, DMOS values and HM data of subjects are also included. & \url{https://github.com/Archer-Tatsu/head-tracking} \\
\hline
Tran \textit{et al.}  \cite{tran2018study} & Video & \tabincell{c}{37\\Ages: 20-37} & 6/126 & \multicolumn{1}{c|}{---} & 30 s & H.265 with different QPs; different resolution & HMD: Samsung Gear VR + Samsung Galaxy S6. The ACR method with 5 quality levels was adopted in the experiment. 18 subjects viewed half of the sequences, and the remained 19 subjects viewed the other half. MOS values are also calculated. & \multicolumn{1}{c|}{N/A} \\
\hline
VQA-ODV \cite{li2018bridge} & Video & \tabincell{c}{221\\Ages: 19-35} & 60/600 & 3840$\times$1920 to 7680$\times$3840 pixels & 10-23 s & H.265 with different QPs; different  projection & HMD: HTC Vive. Eye-tracker: aGlass DKI. The SS method with continuous quality scale was adopted in the experiment. The sequences were equally divided into 10 groups, and each subject only viewed 1 group of sequences (60 sequences in total with 6 reference sequences). Each sequence was viewed by at least 20 subjects. MOS and DMOS values, HM and EM data of subjects are also included. & \url{https://github.com/Archer-Tatsu/VQA-ODV} \\
\hline
\multicolumn{9}{l}{* Data in this column are in the following format: number of reference images or video sequences/number of all images or video sequences.}\\
\multicolumn{9}{l}{\# For videos, this represents the temporal length of the videos themselves; For images, this represents the time each subject view each image in the experiment.}\\
\end{tabular}%
}%
\end{table*}%

\subsubsection{\textbf{Subjective score processing}}
Since raw quality scores rated by subjects are noisy and biased, post-processing is needed to obtain the final quality scores for test materials. A majority of works for subjective VQA on 360{\dg} video/image simply adopted the widely used processed scores in 2D video/image, \textit{i.e.}, mean opinion score (MOS) \cite{itut2006mean} and differential MOS (DMOS) \cite{seshadrinathan2010study}. In contrast, Xu \textit{et al.} \cite{xu2017subjective,xu2018assessing} proposed calculating overall DMOS (O-DMOS) and vectorized DMOS (V-DMOS) for 360{\dg} video. The calculation of O-DMOS is identical to that of DMOS for 2D video/image, which indicates the overall quality of each test sequence. In the calculation of V-DMOS, the HM data of subjects are also used, which reflect the quality of different regions of 360{\dg} sequences. Specifically, the frequency ratio that subject $m$ views region $r$ in sequence $n$ is calculated, denoted as $f_{mn}^{r}$. When $f_{mn}^{r}>f_0$, where $f_0$ is a threshold, subject $m$ is added to collection $\mathbf{M}_n^r$. For region $r$ that $\mathbf{M}_n^r\neq\varnothing$, assuming the re-scaled Z-score \cite{seshadrinathan2010study} of subject $m$ rates for sequence $n$ is $z_{mn}$, the DMOS value  in sequence $n$ can be obtained by
\begin{equation}
  \mathrm{DMOS}_{j}^{r}=\frac{1}{M_n^r}\sum_{m\in\mathbf{M}_n^r} z_{mn}\mbox{,}
\end{equation}
where $M_n^r$ is the size of $\mathbf{M}_n^r$. Finally, the vector of V-DMOS for sequence $n$ can be represented by
\begin{equation}
  [\mathrm{O}\text{-}\mathrm{DMOS}_n \ \; \mathrm{DMOS}_n^1 \ \;\cdots\ \; \mathrm{DMOS}_n^r \ \;\cdots\ \; \mathrm{DMOS}_n^R]\mbox{,}
\end{equation}
where $R$ is the total number of regions in 360{\dg} sequences.

\subsubsection{\textbf{Datasets}}
Based on the subjective experiments, there are many datasets on 360{\dg} video/image for VQA, which include 360{\dg} images or video sequence under different types of distortion with the corresponding subjective quality scores. Table \ref{tab:vqa:ds} summarizes several 360{\dg} video/image datasets for VQA. Among these datasets, the dataset proposed by Upenik \textit{et al.} \cite{upenik2016testbed} in 2016 is one of the earliest VQA dataset for 360{\dg} image. It contains 6 reference images and 54 impaired images. In this dataset, two types of distortion are considered, \textit{i.e.}, the compression distortion of JPEG and projection distortion of ERP and CMP. A total of 48 subjects participated the subjective experiment, who were equally divided into 2 groups. One group of subjects viewed sequences under ERP, while the other group of subjects viewed sequences under CMP. In addition to the MOS values, the HM data of subjects during the experiment are also included.

Zhang \textit{et al.} \cite{zhang2017subjective} proposed a large-scale dataset for 360{\dg} video, in which 16 reference sequences and 384 impaired sequences are included. In this dataset, multiple distortion types are considered: compression distortion of VP9, H.264 and H.265, Gaussian noise and box blur. For each distortion type, there are different levels covering a large quality span, which results in the large gap in the amount between the reference and impaired sequences.

VQA-ODV \cite{li2018bridge} is the largest VQA dataset in Table \ref{tab:vqa:ds} for 360{\dg} video. In VQA-ODV, 540 impaired sequences from 60 reference sequences are included, under the distortions of H.265 at different quantization parameters (QPs) and different projections. A total of 221 subjects participated in the subjective experiment. Therefore, each subject only viewed a subset of 60 sequences and each sequence was viewed by at least 20 subjects. In addition to the MOS and DMOS values for subjective quality, the HM and EM data of subjects are also included in VQA-ODV, enabling the study on the relationship between visual quality and attention of subjects.

Although there have existed several VQA datasets for 360{\dg} video/image, little of the datasets can be downloaded online. Therefore, more VQA datasets for 360{\dg} video/image openly available are urgently needed for evaluation and horizontal comparison of objective VQA approaches.

\subsection{Objective VQA approaches}
In this section, we review the objective VQA approaches for 360{\dg} video/image. For conciseness, we refer to objective VQA approaches by mentioning VQA approaches. It is intuitive to directly apply 2D VQA approaches on 360{\dg} video/image, since 360{\dg} images or video sequences are still encoded and transmitted in the 2D format under sphere-to-plane projection. However, it has been verified that in existing projection types, the sampling density from sphere to plane is not uniform at every pixel locations \cite{yu2015framework}. In this case, directly applying 2D VQA approaches brings bias on the contribution to the quality score of distortion at different pixel locations. Moreover, since VQA approaches predict the subjective quality scores rated by subjects, there have been some 2D VQA approaches introducing perception into VQA \cite{zhang2014vsi,zhang2017study}. As for 360{\dg} video, more than 30\% of the content is not viewed by subjects \cite{li2018bridge}, and thus the distortion in the salient regions of 360{\dg} video/image may contribute more to visual quality. Corresponding to the above two factors, VQA approaches for 360{\dg} video/image can be roughly classified into two categories, one aiming at solving non-uniform sampling and the other incorporating human perception in VQA.

\subsubsection{\textbf{Sampling-related approaches}}
A majority of the sampling-related VQA approaches for 360{\dg} video/image are based on the 2D approaches of peak signal to noise ratio (PSNR) and structural similarity (SSIM) \cite{wang2004image}. Both PSNR and SSIM quantify pixel-wise distortions, and then average the quantified values over all pixel locations to obtain the objective VQA score for 360{\dg} image or video frame. For 360{\dg} video, another averaging over all frames is also implemented. Without loss of generality, both the calculation of PSNR and SSIM can be represented as the following equation ignoring temporal average:
\begin{equation}\label{eq:psnr}
  Q=g\left(\frac{1}{N}\sum_{(i,j)}D(i,j)\right)\mbox{,}
\end{equation}
where $D(i,j)$ is the quantified distortion at the pixel location of $(i,j)$, $N$ is the number of pixels and $g(\cdot)$ is a linear or non-linear function. In the calculation of PSNR, $D(i,j)$ is the squared error between the reference and impaired images/frames of $I$ and $I'$:
\begin{equation}
  D_\text{PSNR}=\left(I(i,j)-I'(i,j)\right)^2\mbox{,}
\end{equation}
and a logarithmic function is then applied as
\begin{equation}
  g_\text{PSNR}(x)=10\log_{10}\left(\frac{I_{\text{max}}^2}{x}\right)\mbox{,}
\end{equation}
where $I_{\text{max}}^2$ is the maximum possible pixel value of the video/image. For video/image with 8-bit precision, $I_{\max}=\num{255}$.
As for SSIM, the distortion is measured in the window located at each pixel location. The similarity index between windows located at the same location in reference and impaired image/frame is measured as the quantification of distortion, from the aspects of luminance, contrast and structure. The function $g(\cdot)$ in SSIM is taken as an identity function, which means that quantified distortion at all pixel locations is directly summed up to be the objective quality score of the video/image.

For solving the non-uniform sampling density of projection, one solution is to re-project the 360{\dg} video/image onto other domains with uniform sampling density. One of the earliest works is the spherical PSNR (S-PSNR) proposed by Yu \textit{et al.} \cite{yu2015framework}, in which PSNR is calculated in the spherical domain. Specifically, $(i,j)$ in \eqref{eq:psnr} is replaced with a set of uniformly distributed points $\mathbb{P}$ on sphere, and $N$ is replaced with the number of these points $N_\text{S-PSNR}$. For each point location $p$ on sphere, its value is obtained at the corresponding pixel location on the original projection plane via the nearest neighbour or bilinear interpolation methods, generating two variants of S-PSNR-NN and S-PSNR-I \cite{boyce2017jvet}. However, in the official implementation of S-PSNR, $N_\text{S-PSNR}=\num{655362}$, which is far less than the number of pixels in typical 360{\dg} video/image with resolution higher than 2K.

Zakharchenko \textit{et al.} proposed an approach calculating PSNR on the projection plane of Craster parabolic projection (CPP) instead of sphere, which is called CPP-PSNR \cite{zakharchenko2016quality,zakharchenko2017omnidirectional}. An illustration of CPP is shown in Figure \ref{fig:sal:proj}. By incorporating CPP, the uniform sampling density is guaranteed, and the resolution of 360{\dg} video/image under CPP is hardly degraded, so that massive information loss is avoided. In CPP-PSNR, only at the pixel locations with actual pixel values is PSNR calculated, \textit{i.e.}, the black regions near the borders are ignored. Correspondingly, $N$ in \eqref{eq:psnr} for CPP-PSNR represents the number of the pixels with actual pixel values. Youvalari \textit{et al.} proposed uniformly sampled spherical PSNR (USS-PSNR) \cite{youvalari2016analysis}, in which PSNR is calculated on the re-sampled 360{\dg} frames. The original 360{\dg} frames are re-sampled according to the circumference of the latitude-wise
slices on sphere. Coincidentally, CPP derives from this re-sampling, such that USS-PSNR is the same as CPP-PSNR.

Although these re-projection approaches completely solve the problem of non-uniform sampling density, the time complexity of these approaches is extremely high, since the procedure of projection is time-consuming \cite{tran2018study}. Therefore, there emerged other approaches incorporating weight allocation in the calculation of PSNR and SSIM to balance the contribution of distortion at different pixel locations. A universal formulation of the approaches with weight allocation can be extended from \eqref{eq:psnr}:
\begin{equation}\label{eq:weight}
  Q=g\left(\frac{\sum_{(i,j)}w(i,j)D(i,j)}{\sum_{(i,j)}w(i,j)}\right)\mbox{,}
\end{equation}
where $w(i,j)$ is the allocated weight at pixel location of $(i,j)$. It is obvious that \eqref{eq:psnr} is a special case of \eqref{eq:weight} when $w(i,j)$ always equals to a constant, implying that all pixels are considered equally important in the origin calculation of PSNR and SSIM. However, in 360{\dg} video/image, distortion at pixel locations with large sampling density is over-weighted, and vice versa.

To solve the overweight problem, Sun \textit{et al.} \cite{sun2017weighted} calculated the stretching ratio in continuous spatial domain for different types of projection, and then the weight maps were derived for these projections to balance the non-uniform sampling density. The weighted-to-spherically-uniform PSNR (WS-PSNR) is proposed to calculated the weighted PSNR with these weight maps. Taking ERP as an example, the weight map in WS-PSNR can be represented as
\begin{equation}\label{eq:wspsnr}
  w_\text{WS-PSNR}^\text{ERP}(i,j)=\cos\left(\phi(j)\right)\mbox{,}
\end{equation}
where $\phi(j)$ is the corresponding latitude of the ordinate $j$ in the pixel coordinate of ERP.
Xiu \textit{et al.} \cite{xiu2017evaluation} derived the weight map from the area that each pixel cover on the sphere, proposing an area weighted spherical PSNR (AW-SPSNR). In AW-SPSNR, distortion is allocated large weight at pixel locations covering large area on the sphere, and vice versa. The weight map in AW-SPSNR can be formulated as
\begin{equation}\label{eq:awspsnr}
  w_\text{AW-SPSNR}(i,j)=\cos\left(\phi(i,j)\right)\cdot|d\theta|_{(i,j)}\cdot|d\phi|_{(i,j)}\mbox{,}
\end{equation}
where $\theta(i,j)$ and $\phi(i,j)$ is the corresponding longitude and latitude of pixel $(i,j)$. For ERP, \eqref{eq:awspsnr} is equivalent to \eqref{eq:wspsnr}.

Instead of calculating PSNR, Lopes \textit{et al.} proposed calculating SSIM and multi-scale SSIM (MS-SSIM) \cite{wang2003multiscale} on the ERP for 360{\dg} video/image. The proposed weighted SSIM (W-SSIM) and weighted MS-SSIM (WMS-SSIM) are calculated with weight allocation, in which the same weight map as \eqref{eq:wspsnr} is adopted.
Similarly, Zhou \textit{et al.} proposed weighted-to-spherically-uniform SSIM (WS-SSIM) \cite{zhou2018weighted}, in which the same weight maps as those in WS-PSNR \cite{sun2017weighted} are applied in the calculation of SSIM.
However, the WS-SSIM, W-SSIM and WMS-SSIM directly calculate the similarity between windows on the 2D projection plane, which is affected by the geometric distortion brought by projection. In contrast, Chen \textit{et al.} proposed a spherical SSIM (S-SSIM) \cite{chen2018spherical} approach, which calculates the similarity between windows on the sphere. To calculate the similarity on windows at each pixel location, small viewport images centered at the spherical location corresponding to the pixel are extracted for both reference and impaired 360{\dg} image/frame. Similarity is calculated between the two extracted viewport images. In S-SSIM, the weight allocation method is the same as \cite{sun2017weighted}. Therefore, both re-projection and weight allocation are incorporated in S-SSIM.
\subsubsection{\textbf{Perceptual approaches}}
For VQA on 2D video/image with perception, an intuitive and simple way is to apply weight allocation in existing VQA approaches, using predicted saliency maps as the weight maps \cite{ma2008image,ma2008saliency}. This idea is also applicable to VQA on 360{\dg} video/image, which is based on attention models discussed in Section \ref{sec:sal}. Yu \textit{et al.} \cite{yu2015framework} proposed two variants of their S-PSNR approach for 360{\dg} video. Two statistical distributions of HM data of subjects are obtained along with only latitude and both longitude and latitude. Then, latitude-wise and pixel-wise weight maps are generated from the two distributions for weight allocation in the calculation of S-PSNR, in the manner of \eqref{eq:weight}. Similarly, Xu \textit{et al.} \cite{xu2018assessing} proposed a non-content-based perceptual PSNR (NCP-PSNR) approach for 360{\dg} video, in which distribution of HM data along longitude and latitude is also obtained on the VR-HM48 dataset to generate a weight map of HM for weight allocation. Considering that the distribution of HM actually represents the distribution of viewport regions, the HM weight map is further convolved by the viewport region to generate the non-content-based weight map, which is utilized to allocate weight in PSNR to calculate the NCP-PSNR.

The above approaches utilized the statistical distribution of HM as the perception cue in weight allocation. Although this kind of distribution is universal, it fails to model attention of subjects well for specific 360{\dg} images and video sequences. Therefore, other approaches proposed predicting attention of subjects and then generating dynamic weight maps for different 360{\dg} images or video sequences, for calculating the weighted PSNR. Luz \textit{et al.} proposed a saliency biased PSNR (SAL-PSNR) \cite{luz2017saliency} for 360{\dg} image, also allocating weights in the calculation of PSNR. The predicted saliency maps from their SDM approach are utilized as the weight maps, multiplied by a distortion factor map that represents the geometric distortion of projection. Xu \textit{et al.} \cite{xu2018assessing} proposed a content-based perceptual PSNR (CP-PSNR) approach for 360{\dg} video, calculating weighted PSNR, too. In CP-PSNR, a random forest model is trained on the VR-HM48 dataset, to predict the most attracted viewport region for each 360{\dg} video frame, based on the previous viewport region. Then, the non-content-based weight map used in the NCP-PSNR is masked by the predicted viewport region, as the content-based weight map for calculating weighted PSNR, called CP-PSNR.
In addition, visual attention spherical weighted PSNR (VASW-PSNR) \cite{ozcinar2019visual}  directly uses the ground-truth saliency maps to weigh pixel-wise distortion, while viewport PSNR (V-PSNR) \cite{yu2015framework} only calculates PSNR in the actual viewports of subjects.

Apart from the approaches that utilize attention models in calculating weighted PSNR, there are other approaches that directly extract perceptual features for VQA on 360{\dg} video/image. Yang \textit{et al.} \cite{yang2017objective} proposed extracting multi-level perceptual features for VQA on 360{\dg} video, including pixel-level \cite{murray2011saliency} and super-pixel-level \cite{yang2013saliency} saliency maps, semantic segmentation \cite{shelhamer2017fully} and equator bias. After temporal transformation, linear regression (LR) and back propagation (BP) neural network are utilized to fuse the features together for obtaining the quality score, namely the LR- and BP-based quality assessment of 360{\dg} videos in the VR system (LR- and BP-QAVR). Huang \textit{et al.} \cite{huang2018modeling} found that the subjective visual quality is related to the spatial resolution and quality factor (QF) of JPEG compression of 360{\dg} image. Therefore, they proposed modelling these two factors via the inverted falling exponential function \cite{ou2011perceptual} with modification. Given the spatial resolution $N$ and QF $q$, the quality score can be modelled by
\begin{equation}\label{eq:sqf}
  Q(N,q)=Q_{\text{max}}\frac{1-\mathrm{e}^{-a_1\left(\frac{N}{N_{\text{max}}}\right)^{0.7}}}{1-\mathrm{e}^{-a_1}} \frac{1-\mathrm{e}^{-b(N)\left(\frac{q}{q_{\text{max}}}\right)}}{1-\mathrm{e}^{-b(N)}}\mbox{,}
\end{equation}
where
\begin{align}
N_{\text{max}}&=4096\times2160,&q_{\text{max}}&=100\mbox{,}\\
b(N)&=a_2\left(\frac{N}{N_{\text{max}}}\right)+a_3,&Q_{\text{max}}&=Q(N_{\text{max}},q_{\text{max}})\mbox{.}
\end{align}
In the above equations, $a_1$, $a_2$, $a_3$ are learnable parameters optimized on the dataset of \cite{huang2018modeling} with the least square error (LSE) criterion.

Given the extraordinary performance in feature extraction of DNN, there have emerged some DNN-based approaches proposed for VQA on 360{\dg} video/image, which also take human perception into account. Kim \textit{et al.} \cite{lim2018vr,kim2019deep} proposed a CNN-based approach with adversarial learning, named VR image quality assessment deep learning framework (DeepVR-IQA). There are a predictor and a guider module in DeepVR-IQA. In the predictor module, the input impaired 360{\dg} image is cropped into patches, and the quality score for each patch is predicted via extracting a quality-related feature vector with a trained CNN. The weight for each patch is also predicted via training another CNN, by extracting the positional feature vector for the patch and combining it with the extracted quality-related feature vector. The quality score of the input 360{\dg} image is obtained by summing up the weighted quality scores of all patches. In the guider module, given the input of reference and impaired 360{\dg} image as well as the predicted and ground-truth quality score, a CNN is trained to distinguish the predicted quality score and the score rated by human. Through adversarial learning, the guider module acts as the supervision of the predictor module. Finally, the visual quality of 360{\dg} image can be effectively rated by predicting the subjective scores of human.
 
 Li \textit{et al.} \cite{li2018bridge} proposed a CNN model for VQA on 360{\dg} video, in which predicted HM and EM are embedded in the CNN structure. Specifically, the HM and EM maps of the input 360{\dg} video sequence are predicted with the visual attention models of DHP \cite{xu2018predicting} and SalGAN \cite{pan2017salgan}. 
Then, 360{\dg} video is cropped into patches. 
Patches are sampled with the corresponding predicted HM values, \textit{i.e.}, patches with low probability of being viewed are discarded. The quality score of each patch is predicted with a trained CNN. Then, these quality scores are weighted with the corresponding predicted EM values of the patches. The weighted quality scores are summed up to obtain the final quality score of the input 360{\dg} video, followed by two fully-connected layers modelling a non-linear function.

\begin{table*}[t]
  \centering
  \caption{Summary of VQA approaches for 360{\dg} video/image.}\label{tab:vqa:approach}%
  \resizebox{\textwidth}{!}{%
\begin{tabular}{|c|c|c|c|c|c|c|c|}
\hline
\multicolumn{2}{|c|}{Approach} & Year & Image/Video & Re-projection & Weight allocation & Data-driven & Availability\\
\hline
\multirow{7}[5]{*}{Sampling-related} & S-PSNR \cite{yu2015framework} & 2015 & Video & \checkmark &   &   & Yes\\
\cline{2-8}  & \tabincell{c}{CPP-PSNR \cite{zakharchenko2016quality,zakharchenko2017omnidirectional}\\USS-PSNR \cite{youvalari2016analysis}} & 2016 & Video & \checkmark &   &   & Yes\\
\cline{2-8}  & WS-PSNR \cite{sun2017weighted} & 2017 & Video &   & \checkmark &   & Yes\\
\cline{2-8}  & AW-SPSNR \cite{xiu2017evaluation} & 2017 & Video &   & \checkmark &   & No\\
\cline{2-8}  & W-SSIM, WMS-SSIM \cite{lopes2018subjective} & 2017 & Video &   & \checkmark &   & No\\
\cline{2-8} & WS-SSIM \cite{zhou2018weighted} & 2018 & Video &   & \checkmark &   & No\\
\cline{2-8}  & S-SSIM \cite{chen2018spherical} & 2018 & Video & \checkmark & \checkmark &   & No\\
\hline
\multirow{10}{*}{Perceptual} & \tabincell{c}{Latitude- \& sphere-weighted S-PSNR\\ (lwS-PSNR \& swS-PSNR) \cite{yu2015framework}} & 2015 & Video & \checkmark & \checkmark & \checkmark & Yes\\
\cline{2-8}  & NCP-PSNR, CP-PSNR \cite{xu2018assessing} & 2018 & Video &   & \checkmark & \checkmark & Yes\\
\cline{2-8}  & SAL-PSNR \cite{luz2017saliency} & 2017 & Image &   & \checkmark & \checkmark & No\\
\cline{2-8} & VASW-PSNR \cite{ozcinar2019visual} & 2019 & Video &   & \checkmark &  \checkmark & No\\
\cline{2-8} & V-PSNR \cite{yu2015framework} & 2015 & Video & \checkmark &   & \checkmark  & No\\
\cline{2-8}  & LR-QAVR, BP-QAVR \cite{yang2017objective} & 2017 & Video &   &   & \checkmark & No\\
\cline{2-8}  & Huang \textit{et al.} \cite{huang2018modeling} & 2018 & Image &   &   & \checkmark & No\\
\cline{2-8}  & DeepVR-IQA \cite{lim2018vr,kim2019deep} & 2019 & Image &   &   & \checkmark & No\\
\cline{2-8}  & Li \textit{et al.} \cite{li2018bridge} & 2018 & Video &   &   & \checkmark & No\\
\cline{2-8}  & V-CNN \cite{li2019viewport} & 2019 & Video & \checkmark &   & \checkmark & Yes\\
\hline
\end{tabular}%
}
\end{table*}%
The above approaches are based on patches, which are generally adopted in DNN-based approaches for 2D video/image \cite{bosse2018deep,pan2018blind,lin2018hallucinated,kim2018deep}. Considering that subjects see viewports rather than patches, Li \textit{et al.} proposed a viewport-based CNN (V-CNN) approach \cite{li2019viewport} for VQA on 360{\dg} video. In addition to the main task of VQA, V-CNN has two stages and several auxiliary tasks. In the first stage, for the input 360{\dg} video frame, several viewport locations with their corresponding weights are proposed by a spherical CNN \cite{cohen2018spherical} model and the viewport softer non maximum suppression (NMS) algorithm. In the second stage, given the input of each proposed viewport image, the saliency map of the viewport is predicted with a mini-DenseNet \cite{huang2017densely}. After multiplying the error map of the viewport with the predicted saliency map, the quality score of the viewport is predicted by another CNN architecture. Finally, the predicted quality score of the input frame is obtained via weighted average over the quality scores of the proposed viewports.

The same as those for 2D video/image, for evaluating the performance of VQA approaches on 360{\dg} video/image, the correlation and error between the objective and subjective quality scores are calculated as the evaluation metrics, including PLCC, Spearman rank-order correlation coefficient (SROCC), Kendall rank-order correlation coefficient (KROCC), root-mean-square error (RMSE) and mean absolute error (MAE). Due to the lack of openly available VQA datasets for 360{\dg} video/image mentioned in Section \ref{sec:vqa:ds}, almost all authors built their own datasets to evaluate the performance of their approaches. Therefore, a benchmark with a large-scale is in urgent need. Table \ref{tab:vqa:approach} summarizes the above reviewed approaches for VQA on 360{\dg} video/image.

\section{Compression}

In order to provide immersive experience and high visual quality for the viewers, 4K or even higher resolution is required for 360{\dg} video/image. Therefore, it is crucial to develop efficient compression approaches to satisfy the need of storage and transmission for 360{\dg} video/image. Generally, 360{\dg} images and video sequences are usually projected to 2D planes to utilize the existing traditional video/image compression standards \cite{chen2018recent}. However, projection may also introduce two problems. First, projected 2D planes suffer from geometric distortion \cite{wang2017new,ghaznavi2018geometry} and redundancy \cite{youvalari2016analysis} caused by non-uniform sampling density. Second, unlike traditional 2D video/image, 360{\dg} video/image is border-less due to its spherical characteristics\footnote{Topologically, a sphere is a two-dimensional closed surface embedded in a three-dimensional Euclidean space, which is compact and without boundary; Geometrically, a sphere is one single surface so that there exists no intersection line between surfaces.}. Sphere-to-plane projection adds artificial borders and thus leads to discontinuity \cite{he2017motion,li2019advanced}. Due to these two problems, the traditional video/image compression standards are unsuitable to be implemented directly on projected 2D planes for 360{\dg} video/image. Therefore, many approaches have been proposed to compress 360{\dg} video/image for a better trade-off between compression efficiency and visual quality.

In the following sections, we classify and review the 360{\dg} video/image compression approaches from the following aspects: motion estimation (ME) adaptation, sampling density correction, re-projection and perceptual compression. 

\subsection{Motion estimation adaptation}

ME and the subsequent motion compensation (MC) are important steps in video compression for reducing temporal redundancy across frames, which are adopted by most of video coding standards (H.264, H.265, etc.). The traditional ME is based on the block matching algorithms with two assumptions. The first one is that any motion of an object at one frame can be effectively represented by block translations in the 2D plane and the corresponding motion vectors (MVs). The second one is that if ME uses any samples outside the borders, padding can handle this case by replicating sample values near frame borders. However, neither of these two assumptions remains true for 360{\dg} video. In fact, geometric distortion brings non-linear motions like rotation and zooming \cite{wang2017new}, which is beyond the representation capability of the traditional motion models. Moreover, MV variation is more severe and thus implementing traditional motion models directly increases motion vectors difference (MVD) between blocks, which leads to the increase of bitrate \cite{ghaznavi2018geometry}. Secondly, traditional padding is not appropriate for 360{\dg} video due to the discontinuity caused by projection \cite{he2017motion} and may cause serious quality degradation \cite{li2019advanced}. Therefore, several approaches have been proposed to modify the process of ME in 360{\dg} video coding. 

\subsubsection{\textbf{ME in spherical domain}}

One intuitive idea is to implement ME in the spherical domain rather than the projected 2D plane. Tosic \textit{et al.} \cite{tosic2005multiresolution} proposed a pioneering approach for ME between 360{\dg} video frames. The first step is to generate multi-resolution representations of current blocks and their corresponding reference blocks in the spherical domain. Then, ME is iteratively refined at successive resolution levels. Recently, Li \textit{et al.} \cite{li2017projection} proposed the similar idea for ME in 360{\dg} video. Current blocks and their corresponding reference blocks in the 360{\dg} frame under CMP are first projected back to the sphere. Then, MC is performed considering 3-dimensional (3D) displacement between spherical blocks. Vishwanath \textit{et al.} \cite{vishwanath2017rotational} further advanced the approach of \cite{li2017projection}. Instead of the block displacement of \cite{li2017projection}, rotations of pixels within a block on the sphere are adopted for better representation of motion. Most recently, Li \textit{et al.} \cite{li2019advanced} further extended their approach \cite{li2017projection} to different projection types. Also inspired by \cite{tosic2005multiresolution}, Simone \textit{et al.} \cite{de2017deformable} introduced a motion plane that tangent to the projected plane, in which any motion in the 360{\dg} video can be approximated by translation movements on the motion plane. An adapted block matching algorithm was also proposed on the basis of the spherical geometry characteristics of the motion plane. Wang \textit{et al.} proposed a spherical coordinate transform-based motion model (SCTMM) \cite{wang2019spherical}, in which the motion of each coding block is considered as 3D translation in the spherical domain. The 3D translation can be represented with a 2D MV and a relative depth parameter, so that it is easy to integrate SCTMM into existing 2D video coding schemes for improving the compression efficiency.

\subsubsection{\textbf{MV mechanism modification}}

Another reasonable idea is to tailor the MV mechanism according to the characteristics of projection formats. Zheng \textit{et al.} \cite{zheng2007adaptive} first proposed a motion-compensated prediction scheme for 360{\dg} video with two phases to extract MVs representing complex motions in ERP. The first phase is traditional ME, which generates block pairs and their corresponding MVs. In the second phase, the affine motion model followed by the quadratic motion model is performed to further refine MVs. Recently, Wang \textit{et al.} \cite{wang2017new} proposed enhancing the representation ability of MV mechanism implemented on 360{\dg} video under ERP. Originally, MVs of all pixels within a block are identical. In the approach of \cite{wang2017new}, pixels in a block are allowed to be assigned to different MVs. Taking the characteristics of ERP into account, pixel-wise MVs from the block-level MV are computed using spherical coordinates and relative depth estimation. Youvalari \textit{et al.} \cite{ghaznavi2018geometry} introduced geometry-based MV scaling to mitigate divergence of MVs for 360{\dg} video encoding. To be more specific, according to WS-PSNR \cite{sun2017weighted}, each pixel in ERP plane is allocated a weight to simulate motion behavior in the sphere. After that, all original MVs are re-scaled by their scaling factors (calculated according to weights). Consequently, a more uniform MV distribution is achieved. 

\subsubsection{\textbf{Padding method adaptation}}
It has been extensively studied to modify the padding method for considering discontinuity.
Sauer \textit{et al.} \cite{sauer2017improved} proposed a cube face extension approach for 360{\dg} video under CMP. Specifically, geometric distortion between cube faces is first corrected based on camera calibration.
Then, each cube face is padded by 4 isosceles trapezoid extensions outside the 4 borders. In contrast, He \textit{et al.} \cite{he2017motion} proposed implementing geometry padding for 360{\dg} video under ERP and CMP, respectively. Unlike conventional padding that simply performs pixel repetition \cite{sullivan2012overview}, geometry padding \cite{he2017motion} for ERP is based on spherical projection to enhance continuity of neighboring samples and contain more meaningful information from the spherical source. For CMP, the proposed geometry padding is applied on each of 6 cube faces. Inspired by \cite{sauer2017improved,he2017motion}, Li \textit{et al.} \cite{li2019advanced} proposed local 3D padding that takes cyclic characteristics of ERP into account. Specifically, for 360{\dg} video under ERP, the left border is padded with pixels near the right border and vice versa. In addition, Li \textit{et al.} \cite{li2018reference} considered the encoding direction and location of the block in the 360{\dg} video under ERP. If a block at the right border is being processed, its top-right reference samples are reconstructed with pixels at the left border. 

\subsection{Sampling density correction}

As mentioned above, non-uniform sampling density causes geometric distortion and unnecessary extra bitrate for over-sampled areas in the projected 2D plane. In general, there are two types of solutions, namely down-sampling and adaptive quantization for compression. The first solution is intuitive, which applies smoothing or blurring filters on the redundant regions. The second one embeds sampling density correction in the process of quantization.
Note that the down-sampling solution can be viewed as the pre-processing steps to save bitrates in the subsequent compression procedures.

\subsubsection{\textbf{Down-sampling}}

Budagavi \textit{et al.} \cite{budagavi2015360} first proposed a variable smoothing approach, which applies Gaussian smoothing filter on the top and bottom regions of 360{\dg} video under ERP. Similarly, Youvalari \textit{et al.} \cite{youvalari2016analysis} proposed splitting 360{\dg} video under ERP into several down-sampled strips according to the latitude. Note that in both approaches \cite{budagavi2015360,youvalari2016analysis}, down-sampling becomes more harsh gradually in the two vertical directions towards top and bottom borders. Additionally, Lee \textit{et al.} \cite{lee2017omnidirectional} proposed implementing down-sampling on rhombus-shaped image transformed from 360{\dg} video/image under ERP. 

\subsubsection{\textbf{Adaptive quantization}}

In H.265 coding, QP controls video quality in the corresponding coding unit (CU) as it determines how much distortion can be incurred in the encoding process. Note that if QP is fixed during encoding process on the projected 2D plane, this is equivalent to applying non-uniform quantization to the sphere \cite{xiu2018adaptive}. Recently, the R-$\lambda$ scheme was considered as a state-or-the-art method for rate distortion optimization (RDO) in the encoding process \cite{liu2017novel}. Therefore, how to adjust RDO to 360{\dg} video was investigated. Tang \textit{et al.} \cite{tang2017optimized} proposed adding a multiplier concerning latitude in the original QP formula for 360{\dg} video under ERP, in order to achieve higher quality around the equator regions. Liu \textit{et al.} \cite{liu2017novel,liu2018rate} proposed optimizing 360{\dg} video encoding through maximization on S-PSNR \cite{yu2015framework} at a given bitrate, since S-PSNR is an effective objective VQA approach for 360{\dg} video. Li \textit{et al.} \cite{li2017spherical} proposed that the weight value of WS-PSNR \cite{sun2017weighted} in the center of each block is added as a scalar of Lagrangian multiplier $\lambda$ in RDO. Similarly, Xiu 
\textit{at al.} \cite{xiu2018adaptive} also took WS-PSNR \cite{sun2017weighted} into account in RDO. In \cite{xiu2018adaptive}, the QP values for luma and chroma components are calculated independently, due to their difference of dynamic range.

\subsection{Re-projection approaches}
Recently, 360{\dg} video/image compression has benefited from more compression-friendly re-projection approaches, which can be classified into two categories. One category is to perform modification based on the existing projection types with rotation in spherical domain, in order to find the most compressible spherical position.
The other one is to design novel projection types that take full consideration of spherical characteristics of 360{\dg} video/image and compression for better coherence with both the input and its subsequent coding process.

\subsubsection{\textbf{Rotation in spherical domain}}
Some re-projection approaches proposed rotating the 360{\dg} content in spherical domain, such that the regions requiring high visual quality can be located in the less distorted areas in the projected 2D plane before compression.
Specifically, Boyce \textit{et al.} \cite{boyce2017spherical} proposed 
rotating the regions containing high motion and detailed texture to the front regions near the equator, which corresponds to the center of the ERP frames with less geometric distortion. 
On the other hand, the static regions are rotated to the poles, which correspond to the horizontal borders of the ERP frames. Consequently, those informative regions can incur less distortion. Recently, Su \textit{et al.} \cite{su2018learning} extended the spherical rotation-based re-projection to the CMP frames. As deep learning is introduced to the field of video/image compression \cite{li2018learning,santurkar2018generative}, Su \textit{et al.} \cite{su2018learning} designed a CNN fed by segmentation contours and MVs to learn the association between visual content and its compression rate at different rotations. In \cite{su2018learning}, the optimal rotation for compression can be obtained by the single prediction of the CNN model. 

\subsubsection{\textbf{Compression-friendly projection types}}
Xiu \textit{et al.} \cite{xiu2017evaluation} investigated the impact of the projection types on compression efficiency. It turns out that the choice of the best projection type is highly content-dependent, and quality variation among different types can be significantly large. Based on CMP, Lin \textit{et al.} \cite{lin2019efficient} proposed a modified projection type, called the hybrid equi-angular cubemap (HEC) projection. A traditional equi-angular mapping function and the proposed latitude-related mapping function are applied on different cube faces in HEC to solve the non-uniform sampling density within cube faces of CMP. Based on the characteristic of content-dependence, He \textit{et al.} \cite{he2018content} developed a generalized version of CMP called hybrid cubemap projection (HCP), which allows sampling function selection according to video content.
In addition, some other re-projection approaches utilize the novel projection types that achieve more uniform sampling and continuity, which can also facilitate 360{\dg} video/image compression. Hybrid angular cubemap projection (HAP) proposed by Hanhart \textit{et al.} \cite{hanhart2019360} inherits the content-adaptive property of HCP and further improves compression efficiency by keeping sampling continuity at borders that connect two faces. Li \textit{et al.} \cite{li2016novel} proposed a tile-based segmentation and projection type, minimizing the pixel area as the quantity of tiles increases. Consequently, redundancy among faces can be mitigated for speeding up compression. Wu \textit{et al.} \cite{chengjia2018octagonal} proposed a novel octagonal mapping (OGM) type and the corresponding pixel rearrangement scheme, and it achieves more uniform sampling without generating more seams compared with the traditional projection types. As a result, both geometric distortion and compression complexity can be reduced.

\subsection{Perceptual compression}
360{\dg} video/image compression can take advantage of human perception from two aspects. First, viewers can only see viewports through the HMD \cite{sreedhar2016viewport,taghavinasrabadi2017adaptive}, which merely occupies approximately $12.5\%$ of the whole streaming data \cite{ozcinar2017viewport}. Thus, it is not only a waste of bandwidth but also an unnecessary burden to encode and transmit the whole 360{\dg} video/image with full resolution. The perceptual quality of 360{\dg} video/image mainly depends on the quality within the viewport. Consequently, it is intuitive to allocate more bitrate to the scene in the viewport than other regions. Second, the human visual system (HVS) indicates that viewers tend to pay attention to RoI inside the viewport. Therefore, RoIs deserve higher quality than other regions \cite{luz2017saliency}. On one hand, it is natural that saliency prediction approaches can be applied in intra-frame compression. On the other hand, as objects with high motion in salient regions are most likely to attract viewers' attention, frame rate can be optimized in inter-frame compression, according to the principle that frames containing dynamic salient regions deserve higher frame rate. 
\begin{table*}[t]
  \centering
  \caption{Summary of  360{\dg} video/image compression approaches.}
  \resizebox{\textwidth}{!}{%
\begin{tabular}{|c|c|c|c|c|c|c|c|c|c|c|c|c|c|}
\hline
\multicolumn{2}{|c|}{Approaches} & Year & Image/Video & Codec & Pre-processing & ME\&MC & Quantization & Rate Control & Streaming & Projection Type & Evaluation Method * & VQA Metric & Benchmark \#\\
\hline
\multirow{12}[20]{*}{\rotatebox{90}{\tabincell{c}{Motion estimation\\adaptation}}} & Tosic \textit{et al.} \cite{tosic2005multiresolution} & 2005 & Image & --- & \checkmark & \checkmark &   &   &   & --- & Others & --- & Self-defined\\
\cline{2-14}  & Li \textit{et al.} \cite{li2017projection} & 2017 & Video & H.265 &   & \checkmark &   &   &   & CMP & BD-BR & --- & \tabincell{c}{JVET \cite{boyce2017jvet},\\ GoPro \cite{gopro}}\\
\cline{2-14}  & Vishwanath \textit{et al.} \cite{vishwanath2017rotational} & 2017 & Video & H.265 &   & \checkmark &   &   &   & ERP, CMP & BD-BR & WS-PSNR \cite{sun2017weighted} & JVET \cite{boyce2017jvet}\\
\cline{2-14}  & Li \textit{et al.} \cite{li2019advanced} & 2019 & Video & H.265 &   & \checkmark &   &   &   & \tabincell{c}{ERP, CMP,\\OHP} & BD-BR & \tabincell{c}{WS-PSNR \cite{sun2017weighted}, S-PSNR \cite{yu2015framework},\\CPP-PSNR \cite{zakharchenko2016quality}} & JVET \cite{boyce2017jvet}\\
\cline{2-14}  & Simone \textit{et al.} \cite{de2017deformable} & 2017 & Video & --- &   & \checkmark &   &   &   & ERP & Others & PSNR, SSIM \cite{wang2004image}, S-PSNR \cite{yu2015framework} & Self-defined\\
\cline{2-14}  & SCTMM \cite{wang2019spherical} & 2019 & Video & H.266 &   & \checkmark &   &   &   & ERP & BD-BR, complexity & WS-PSNR \cite{sun2017weighted} & \tabincell{c}{JVET \cite{boyce2017jvet},\\AVS \cite{vrsequences}}\\
\cline{2-14}  & Zheng \textit{et al.} \cite{zheng2007adaptive} & 2007 & Video &  H.264 &   & \checkmark &   &   &   & --- & RD curve, others & PSNR & Self-defined\\
\cline{2-14}  & Wang \textit{et al.} \cite{wang2017new} & 2017 & Video & H.265 &   & \checkmark &   &   &   & ERP & BD-BR & PSNR, SSIM \cite{wang2004image} & AVS \cite{vrsequences}\\
\cline{2-14}  & Youvalari \textit{et al.} \cite{ghaznavi2018geometry} & 2018 & Video & H.266 &   & \checkmark &   &   &   & ERP & BD-BR, complexity & \tabincell{c}{PSNR, WS-PSNR \cite{sun2017weighted},\\CPP-PSNR \cite{zakharchenko2016quality}} & JVET \cite{boyce2017jvet}\\
\cline{2-14}  & Sauer \textit{et al.} \cite{sauer2017improved} & 2017 & Video & H.265 & \checkmark & \checkmark &   &   &   & CMP & BD-BR, BD-PSNR & PSNR & \tabincell{c}{JVET \cite{boyce2017jvet},\\GoPro \cite{gopro}}\\
\cline{2-14}  & He \textit{et al.} \cite{he2017motion} & 2017 & Video & H.265 & \checkmark & \checkmark &   &   &   & ERP, CMP & BD-BR & S-PSNR \cite{yu2015framework} & \tabincell{c}{JVET \cite{boyce2017jvet},\\GoPro \cite{gopro}}\\
\cline{2-14}  & Li \textit{et al.} \cite{li2018reference} & 2018 & Video & H.265 & \checkmark & \checkmark &   &   &   & ERP & BD-BR & PSNR & JVET \cite{boyce2017jvet}\\
\hline
\multirow{7}[14]{*}{\rotatebox{90}{\tabincell{c}{Sampling density\\correction}}} & Budagavi \textit{et al.} \cite{budagavi2015360} & 2015 & Video &  H.264 & \checkmark &   &   &   &   & ERP & BD-BR & --- & Self-defined\\
\cline{2-14}  & Youvalari \textit{et al.} \cite{youvalari2016analysis} & 2016 & Video & H.265 & \checkmark &   &   &   &   & ERP & BD-BR & \tabincell{c}{S-PSNR, lwS-PSNR \cite{yu2015framework},\\USS-PSNR \cite{youvalari2016analysis}} & Self-defined\\
\cline{2-14}  & Lee \textit{et al.} \cite{lee2017omnidirectional} & 2017 & Video & H.265 & \checkmark &   &   &   &   & ERP & BD-BR & S-PSNR, lwS-PSNR \cite{yu2015framework} & \tabincell{c}{SUN360 \cite{xiao2012recognizing},\\self-defined}\\
\cline{2-14}  & Tang \textit{et al.} \cite{tang2017optimized} & 2017 & Video & H.265 &   &   & \checkmark &   &   & ERP & BD-BR & S-PSNR \cite{yu2015framework} & AVS \cite{vrsequences}\\
\cline{2-14}  & Liu \textit{et al.} \cite{liu2017novel} & 2017 & Video & H.265 &   &   & \checkmark & \checkmark &   & ERP & \tabincell{c}{RD curve, BD-BR,\\BD-PSNR, others} & S-PSNR \cite{yu2015framework} & AVS \cite{vrsequences}\\
\cline{2-14}  &  Li \textit{et al.} \cite{li2017spherical} & 2017 & Video & H.265 &   &   & \checkmark &   &   & ERP & RD curve, BD-BR & \tabincell{c}{WS-PSNR \cite{sun2017weighted}, S-PSNR \cite{yu2015framework},\\CPP-PSNR \cite{zakharchenko2016quality}} & JVET \cite{boyce2017jvet}\\
\cline{2-14}  &  Xiu \textit{et al.} \cite{xiu2018adaptive} & 2018 & Video & H.265 &   &   & \checkmark &   &   & ERP, CMP & RD curve, BD-BR & S-PSNR \cite{yu2015framework}, WS-PSNR \cite{sun2017weighted} & JVET \cite{boyce2017jvet}\\
\hline
\multirow{7}[12]{*}{\rotatebox{90}{Re-projection}} & Boyce \textit{et al.} \cite{boyce2017spherical} & 2017 & Video & H.265, H.266 & \checkmark &   &   &   &   & ERP & RD curve, BD-BR & \tabincell{c}{WS-PSNR \cite{sun2017weighted}, S-PSNR \cite{yu2015framework},\\CPP-PSNR \cite{zakharchenko2016quality}} & JVET \cite{boyce2017jvet}\\
\cline{2-14}  & Su \textit{et al.} \cite{su2018learning} & 2018 & Video & \tabincell{c}{H.264, H.265,\\VP9} & \checkmark &   &   &   &   & CMP & Others & --- & Self-defined\\
\cline{2-14}  & HEC \cite{lin2019efficient} & 2019 & Video & H.265, H.266 & \checkmark &   &   &   &   & HEC \cite{lin2019efficient} & RD curve, BD-BR & WS-PSNR \cite{sun2017weighted} & JVET \cite{boyce2017jvet}\\
\cline{2-14}  & HCP \cite{he2018content} & 2018 & Video & H.265 & \checkmark &   &   &   &   & HCP \cite{he2018content} & \tabincell{c}{BD-BR, BD-PSNR,\\complexity} & WS-PSNR \cite{sun2017weighted} & JVET \cite{boyce2017jvet}\\
\cline{2-14}  & HAP \cite{hanhart2019360} & 2019 & Video & H.265, H.266 & \checkmark &   &   &   &   & HAP \cite{hanhart2019360} & BD-BR, complexity & WS-PSNR \cite{sun2017weighted} & JVET \cite{boyce2017jvet}\\
\cline{2-14}  & Li \textit{et al.} \cite{li2016novel} & 2016 & Video & H.265 & \checkmark &   &   &   &   & Li \textit{et al.} \cite{li2016novel} & RD curve, BD-BR & S-PSNR, lwS-PSNR \cite{yu2015framework} & \tabincell{c}{SUN360 \cite{xiao2012recognizing},\\LetinVR \cite{letinvr}}\\
\cline{2-14}  & OGM \cite{chengjia2018octagonal} & 2018 & Video & H.265 & \checkmark &   &   &   &   & OGM \cite{chengjia2018octagonal} & RD curve, BD-BR & WS-PSNR \cite{sun2017weighted}, S-PSNR \cite{yu2015framework} & AVS \cite{vrsequences}\\
\hline
\multirow{12}[15]{*}{\rotatebox{90}{Perceptual}} & Sreedhar \textit{et al.} \cite{sreedhar2016viewport} & 2016 & Video & H.265 &   &   &   &   & \checkmark & \textit{Multiple} & RD curve, BD-BR & PSNR & Self-defined\\
\cline{2-14}  & Corbillon \textit{et al.} \cite{corbillon2017viewport} & 2017 & Video & H.265 &   &   &   &   & \checkmark & ERP & Others & PSNR, MS-SSIM \cite{wang2003multiscale} & Jaunt Inc.\\
\cline{2-14}  & Ozcinar \textit{et al.} \cite{ozcinar2017viewport} & 2017 & Video & H.264 &   &   &   &   & \checkmark & ERP & Others & PSNR, SSIM \cite{wang2004image} & MPEG \cite{bang2016description}\\
\cline{2-14}  & Ozcinar \textit{et al.} \cite{ozcinar2019visual} & 2019 & Video & H.265 &   &   &   &   & \checkmark & ERP & \tabincell{c}{RD curve, BD-PSNR,\\others} & \tabincell{c}{WS-PSNR \cite{sun2017weighted},\\VASW-PSNR \cite{ozcinar2019visual}} & \tabincell{c}{JVET \cite{boyce2017jvet},\\MPEG \cite{bang2016description}}\\
\cline{2-14}  & Nguyen \textit{et al.} \cite{nguyen2019optimal} & 2019 & Video & H.265 &   &   &   &   & \checkmark & ERP & Others & V-PSNR \cite{yu2015framework} & Self-defined\\
\cline{2-14}  & Fuente \textit{et al.} \cite{sanchez2019delay} & 2019 & Video & H.265 &   &   &   &   & \checkmark & CMP & BD-BR, others & PSNR & Self-defined\\
\cline{2-14}  & Nasrabadi \textit{et al.} \cite{taghavinasrabadi2017adaptive,nasrabadi2017adaptive} & 2017 & Video & H.265 &   &   &   &   & \checkmark & CMP & RD curve, BD-BR, others & PSNR & Self-defined\\
\cline{2-14}  & Ozcinar \textit{et al.} \cite{ozcinar2017estimation} & 2017 & Video & H.264 &   &   &   &   & \checkmark & ERP & \tabincell{c}{RD curve, BD-BR,\\complexity} & WS-PSNR \cite{sun2017weighted} & \tabincell{c}{JVET \cite{boyce2017jvet},\\MPEG \cite{bang2016description}}\\
\cline{2-14}  & Sun \textit{et al.} \cite{sun2019two} & 2019 & Video & H.265 &   &   &   &   & \checkmark & ERP & RD curve, others & WS-PSNR \cite{sun2017weighted} & JVET \cite{boyce2017jvet}\\
\cline{2-14}  & Luz \textit{et al.} \cite{luz2017saliency} & 2017 & Image & H.265 &   &   & \checkmark &   &   & ERP & RD curve, BD-BR, others & \tabincell{c}{WS-PSNR \cite{sun2017weighted}, V-PSNR \cite{yu2015framework},\\SAL-PSNR \cite{luz2017saliency}} & Salient360 \cite{rai2017dataset}\\
\cline{2-14}  & Liu \textit{et al.} \cite{liu2018rate} & 2018 & Video & H.265 &   &   & \checkmark & \checkmark &   & ERP & \tabincell{c}{RD curve, BD-BR,\\BD-PSNR, others} & S-PSNR \cite{yu2015framework} & AVS \cite{vrsequences}\\
\cline{2-14}  & Tang \textit{et al.} \cite{tang2017optimized} & 2017 & Video & H.265 & \checkmark &   &   &   &   & CMP & BD-BR & S-PSNR \cite{yu2015framework} & AVS \cite{vrsequences}\\
\hline
\multicolumn{14}{l}{* In this column, we only focus on coding-related evaluation methods: Rate-distortion (RD) curve, Bjøntegaard delta bitrate saving (BD-BR),  Bjøntegaard delta PSNR improvement (BD-PSNR) and complexity.}\\
\multicolumn{14}{l}{\# Due to the different versions of the documents in this column, the test sequences can be different despite on the same benchmark.} \\
\end{tabular}%
}
  \label{tab:compression}%
\end{table*}%
\subsubsection{\textbf{Viewport-based coding/streaming}}
The viewport-based approaches, especially streaming-aware coding approaches, mainly improve compression efficiency by allocating different encoding quality according to the HM of the viewer. They can be classified into two categories: non-tile coding and tile-based coding \cite{ozcinar2017viewport}. 

In the non-tile coding/streaming, 360{\dg} video/image is first projected and packed into the same frame. Then, more bitrate assgined to the primary viewport than the remaining regions \cite{ghaznavi2017comparison}. Sreedhar \textit{et al.} \cite{sreedhar2016viewport} first proposed a streaming approach to transmit the front viewport with high resolution and the remaining parts with relatively lower resolution. Another contribution of \cite{sreedhar2016viewport} is that a streaming-aware HM tracking algorithm is developed for real-time adaptive efficient compression of 360{\dg} video. Likewise, in the recent approach proposed by Corbillon \textit{et al.} \cite{corbillon2017viewport}, visual quality around the HM position is regarded as the maximum, and quality degradation is implemented based on spatial distance from the HM position in the process of streaming. 

The tile-based coding/streaming that independently encodes and transmits each tile outweighs non-tile coding in terms of higher flexibility and less resource consumption \cite{lederer2017today,zare2017virtual}. It is worth mentioning that despite individual processing of each tile, bitrate distribution is still determined by location of the primary viewport in the frame. As a earlier work, \cite{skupin2016tile} proposed performing tile-based coding and streaming on 360{\dg} video. Specifically, video tiles in two different levels of resolution are simultaneously transmitted. Frame reconstruction process is an integration of high resolution tiles within the viewport and low resolution tiles outside of the viewport. 
In recent approaches \cite{ozcinar2017viewport,hosseini2016adaptive,ozcinar2019visual,nguyen2019optimal}, each tile has several hierarchical representation levels to be chosen, depending on the viewport of the viewer. As such, a more smooth quality degradation can be achieved.
Furthermore, Fuente \textit{et al.} \cite{sanchez2019delay} proposed predicting the HM of the viewer from his/her previous HM data, considering both angular velocity and angular acceleration. According to the predicted HM, different QP is allocated to each tile.
Nasrabadi \textit{et al.} \cite{taghavinasrabadi2017adaptive,nasrabadi2017adaptive} inherited the idea of \cite{alface2012interactive} and further proposed a layered 360{\dg} video coding approach, in which each tile is encoded with base quality and several enhancement layers. Another highlight of \cite{taghavinasrabadi2017adaptive} is that given the tiles with base quality, HM is predicted as the guidance of bitrate distribution in future frames.
In addition, inspired by \cite{alface2012interactive,ozcinar2017viewport}, Ozcinar \textit{et al.} \cite{ozcinar2017estimation} proposed encoding ladder estimation, in which the quality of tiles is determined by a trade-off between video content (spatial and temporal features) and network condition. Sun \textit{et al.} \cite{sun2019two} proposed a two-tier system for 360{\dg} video streaming, in which the base tier delivers the entire video content at a lower quality with a long buffer, while the enhancement tier delivers the predicted viewport at a higher quality with a short buffer. Consequently, a trade-off between reliability and the efficiency is achieved for 360{\dg} video streaming.

\subsubsection{\textbf{Saliency-aware adaptive coding}}
Recently, the saliency-aware compression approaches for traditional 2D video/image have been extensively studied \cite{zhu2018innovative}. However, unlike the prosperity of the aforementioned viewport-based approaches, saliency-aware compression for 360{\dg} video/image is still an emerging and challenging topic \cite{biswas2017towards}.
Recently, there have been some pioneering approaches that tune the existing 2D approaches according to the characteristics of 360{\dg} video/image. 

For intra-frame compression of 360{\dg} video/image, saliency can be blended into quantization and sampling process. Recently, Luz \textit{et al.} \cite{luz2017saliency} proposed that the QP values of a 360{\dg} image can be estimated based on its saliency map, which is the output of the proposed SDM reviewed in Section \ref{sec:sal:approach}. To avoid drastic quality difference, the computation of QP is also aware of spatial activity, which contains spatial activity in the luma coding block, mean spatial activity of all CUs etc. Finally, an integrated formula of QP combining saliency and spatial activity is developed. In addition, Sitzmann \textit{et al.} \cite{sitzmann2018saliency} implemented and validated the effectiveness of saliency-based sampling on 360{\dg} video via applying existing saliency prediction approaches. In detail, low-resolution CMP is integrated with up-sampled salient regions for better considering attention preference. Recently, Zhu \textit{et al.} \cite{zhu2018innovative} have proposed a saliency guided RoI selection approach that combines segmentation-based salient objects detection with the saliency prediction model to determine the resolution of regions. Biswas \textit{et al.} \cite{biswas2017towards} further refined the approach of \cite{sitzmann2018saliency} by illumination normalization.

Considering the fact that subject visual quality is mainly influenced by salient regions containing objects with high motions, rate control based on saliency information is a reasonable idea for inter-frame compression of 360{\dg} video. Liu \textit{et al.} \cite{liu2018rate} introduced a rate control scheme for perceptual quality improvement of 360{\dg}
video under ERP. To be more specific, bit allocation is determined by perceptual distortion using front-center-bias saliency map. Similarly, in the approach proposed by Tang \textit{et al.} \cite{tang2017optimized}, if a frame is with low saliency and low motion, this frame can be dropped and replaced by the previous frame encoded at low bitrate. Taking the multi-face characteristics of CMP into account, the frame rate of each face is optimized individually for higher compression efficiency.

\section{Discussion and future work}
\begin{figure}[t]
\begin{center}
\includegraphics[width=\linewidth]{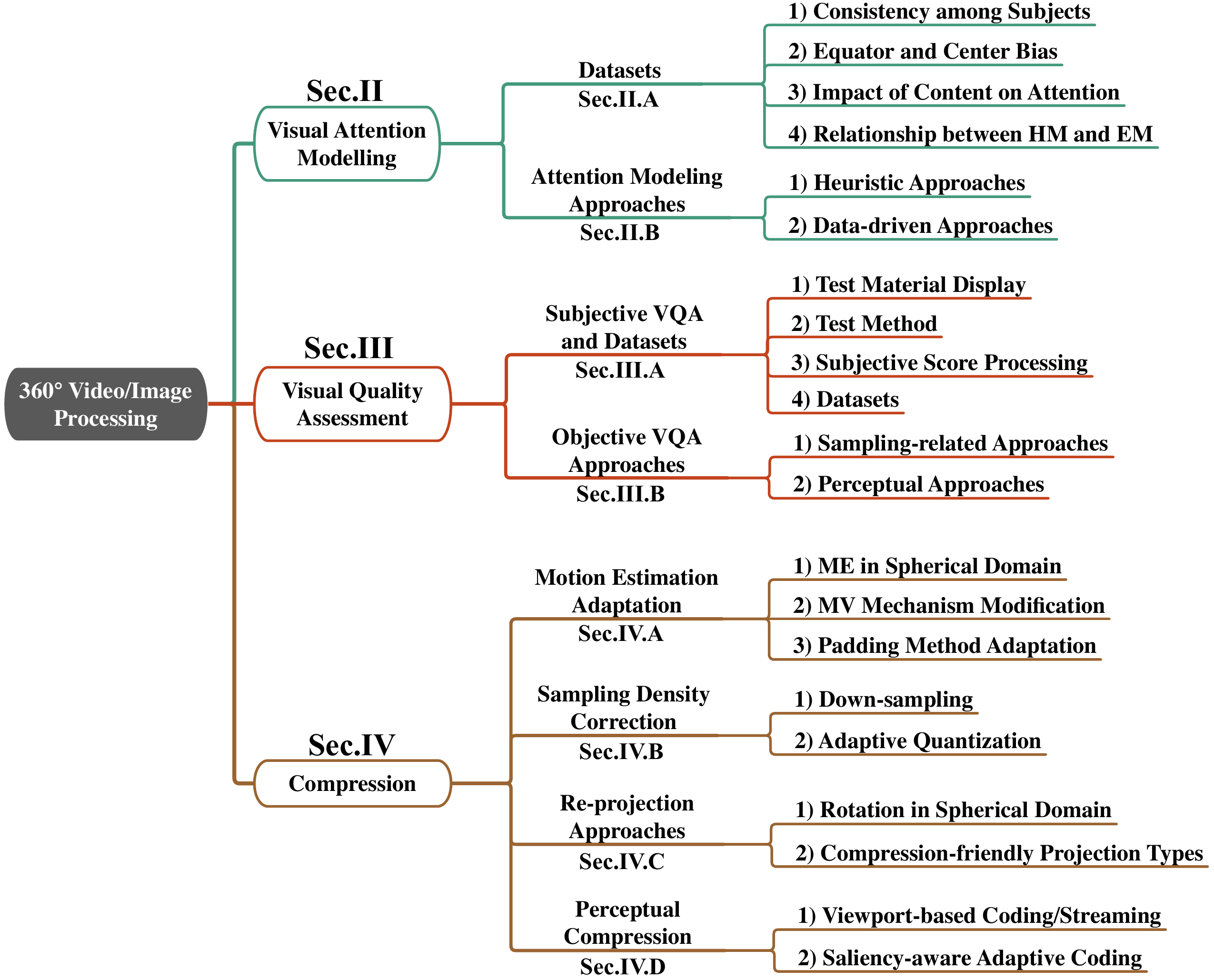}
\end{center}
\caption{Overall organization of this paper.}\label{fig:outline}
\end{figure}

Recently, 360{\dg} video/image has been in great popularity and of broad prospects, which provides immersive experience but meanwhile requires high fidelity and low latency. Consequently, the great challenges are posed to storage and transmission, which calls for dedicated and more efficient processing for 360{\dg} video/image.
In this paper, we surveyed state-of-the-art works on 360{\dg} video/image processing, from the aspects of visual attention modelling, VQA and compression. Figure \ref{fig:outline} illustrates the outline of this paper.

For visual attention modelling, several public 360{\dg} video/image datasets with attention data were overviewed, and based on these datasets, human attention on 360{\dg} video/image was analyzed. Then, we reviewed saliency prediction approaches for 360{\dg} video/image, including both heuristic and data-driven approaches. Heuristic approaches explicitly extract hand-crafted features, so that the interpretability is guaranteed. Data-driven approaches, especially DNN-based approaches, automatically extract features for saliency prediction and achieve considerable performance. However, compared to 2D video/image datasets \cite{borji2018saliency} that include up to \num{4000} images or \num{1000} video sequences, 360{\dg} video/image attention datasets are of rather small size, which may seriously limit the up-bound of the data-driven approaches. Therefore, some large-scale 360{\dg} video/image datasets are in urgent need for attention modelling. It is worth noting that a majority of approaches adopt and modify 2D saliency prediction approaches or extract features that have been verified useful in 2D saliency prediction, such that the well-studied visual attention mechanism on 2D video/image can be utilized. However, there also emerged a handful of approaches \cite{zhang2018saliency,xu2018predicting} developing brand-new mechanism for 360{\dg} video/image. It is reasonable to believe that with more and larger datasets for 360{\dg} video/image being available in the future, visual attention mechanism on 360{\dg} video/image will be revealed in more detail. Consequently, the saliency prediction approaches for 360{\dg} video/image can be further advanced, with increasingly better performance.

Moreover, this paper reviewed both subjective and objective VQA approaches for  360{\dg} video/image. The subjective VQA approaches also provide the datasets with subjective quality scores. Then, the objective VQA approaches aim to predict the subjective quality scores of 360{\dg} video/image.
Several objective VQA approaches solve the non-uniform density caused by projection on 360{\dg} video/image in the calculation of 2D VQA approaches, with re-projection or weight allocation. These sampling-based approaches keep the generialization ability of the wide-spread 2D VQA approaches. Other objective VQA approaches for 360{\dg} video/image incorporate the visual attention models in VQA, in order to make it consistent with subjective quality scores. 
Currently, with the cutting-edge advances made in visual attention models, much more works related to attention-based VQA on 360{\dg} video/image, especially considering viewport, remain to be developed. 
Additionally, most of the existing works focus on extracting spatial features for VQA on 360{\dg} video.
Thus, extracting spatial-temporal features for VQA on 360{\dg} video shows a promising trend in the future.
In addition, since most of the existing 360{\dg} video/image VQA works are full-reference (FR) approaches (except DeepVR-IQA \cite{lim2018vr,kim2019deep}), reduced-reference (RR) or no-reference (NR, \textit{a.k.a}, blind) VQA approaches on 360{\dg} video/image show another promising trend in the future.

For compression, the mainstream is to apply the 2D video coding standards on 360{\dg} video/image with modification, in order to take advantage of well developed 2D video coding standards.
Some approaches take consideration of the spherical characteristics of 360{\dg} video/image in the modification, such as ME adaptation, sampling density correction and re-projection.
Other approaches incorporate visual attention models in compressing 360{\dg} video/image, such that the bitrate can be significantly saved at satisfactory visual quality.
All the existing works are based on the hybrid framework of 2D video coding.
Some brand-new coding frameworks for 360{\dg} video/image may see a great deal of future work.

In fact, the perception, VQA and compression of 360{\dg} video/image do not exist in isolation. For example, the visual attention models may benefit VQA on 360{\dg} video/image, as human is the ultimate end of quality assessment; VQA serves as an optimization goal of 360{\dg} video/image compression. 
Most recently, multi-task learning \cite{argyriou2007multi}, which refers to simultaneously learning more than one tasks, has become mature. In light of multi-task learning, it is possible to simultaneously handle the tasks of perception, VQA and compression for 360{\dg} video/image, as these tasks are not isolated. 
There has been a pioneering work \cite{li2019viewport} that developed a multi-task for predicting viewports, saliency maps and subjective quality scores of 360{\dg} video.
A long-term goal of future 360{\dg} video/image processing should  include more multi-task works for perception, VQA, compression, etc.


\ifCLASSOPTIONcaptionsoff
  \newpage
\fi



\bibliographystyle{IEEEtran}
\bibliography{IEEEfull}

\begin{thebibliography}{100}
\providecommand{\url}[1]{#1}
\csname url@samestyle\endcsname
\providecommand{\newblock}{\relax}
\providecommand{\bibinfo}[2]{#2}
\providecommand{\BIBentrySTDinterwordspacing}{\spaceskip=0pt\relax}
\providecommand{\BIBentryALTinterwordstretchfactor}{4}
\providecommand{\BIBentryALTinterwordspacing}{\spaceskip=\fontdimen2\font plus
\BIBentryALTinterwordstretchfactor\fontdimen3\font minus
  \fontdimen4\font\relax}
\providecommand{\BIBforeignlanguage}[2]{{%
\expandafter\ifx\csname l@#1\endcsname\relax
\typeout{** WARNING: IEEEtran.bst: No hyphenation pattern has been}%
\typeout{** loaded for the language `#1'. Using the pattern for}%
\typeout{** the default language instead.}%
\else
\language=\csname l@#1\endcsname
\fi
#2}}
\providecommand{\BIBdecl}{\relax}
\BIBdecl

\bibitem{chen2018recent}
Z.~Chen, Y.~Li, and Y.~Zhang, ``Recent advances in omnidirectional video coding
  for virtual reality: Projection and evaluation,'' \emph{Signal Processing},
  vol. 146, pp. 66--78, 2018.

\bibitem{wien2019standardization}
M.~Wien, J.~M. Boyce, T.~Stockhammer, and W.-H. Peng, ``Standardization status
  of immersive video coding,'' \emph{{IEEE} Journal on Emerging and Selected
  Topics in Circuits and Systems}, 2019.

\bibitem{jpeg2017call}
{JPEG Requirements Subgroup}, ``Call for proposals on {JPEG} 360 metadata,''
  \emph{{ISO/IEC JTC 1/SC 29/WG 1}}, 2017.

\bibitem{itur2007methodology}
{ITU-R}, ``Methodology for the subjective assessment of video quality in
  multimedia applications,'' \emph{Recommendation {ITU-R BT.1788}}, 2007.

\bibitem{itut2008subjective}
{ITU-T}, ``Subjective video quality assessment methods for multimedia
  applications,'' \emph{Recommendation {ITU-T P.910}}, 2008.

\bibitem{itur2012methodology}
{ITU-R}, ``Methodology for the subjective assessment of the quality of
  television pictures,'' \emph{Recommendation {ITU-R BT.500-13}}, 2012.

\bibitem{seshadrinathan2010study}
K.~Seshadrinathan, R.~Soundararajan, A.~C. Bovik, and L.~K. Cormack, ``Study of
  subjective and objective quality assessment of video,'' \emph{{IEEE}
  Transactions on Image Processing}, vol.~19, no.~6, pp. 1427--1441, 2010.

\bibitem{upenik2016testbed}
E.~Upenik, M.~{\v{R}}e{\v{r}}{\'a}bek, and T.~Ebrahimi, ``Testbed for
  subjective evaluation of omnidirectional visual content,'' in \emph{Picture
  Coding Symposium}.\hskip 1em plus 0.5em minus 0.4em\relax IEEE, 2016, pp.
  1--5.

\bibitem{xu2017subjective}
M.~Xu, C.~Li, Y.~Liu, X.~Deng, and J.~Lu, ``A subjective visual quality
  assessment method of panoramic videos,'' in \emph{{IEEE} International
  Conference on Multimedia and Expo}.\hskip 1em plus 0.5em minus 0.4em\relax
  IEEE, 2017, pp. 517--522.

\bibitem{gutierrez2018toolbox}
J.~Guti{\'e}rrez, E.~David, Y.~Rai, and P.~Le~Callet, ``Toolbox and dataset for
  the development of saliency and scanpath models for omnidirectional/360{\dg}
  still images,'' \emph{Signal Processing: Image Communication}, vol.~69, pp.
  35--42, 2018.

\bibitem{rai2017salient360}
Y.~Rai, P.~Le~Callet, and P.~Guillotel, ``{Salient360!}: Visual attention
  modeling for 360{\dg} images,'' in \emph{{IEEE} International Conference on
  Multimedia and Expo 2017 Grand Challenge}.\hskip 1em plus 0.5em minus
  0.4em\relax IEEE, 2017.

\bibitem{gutierrez2018salient360}
J.~Guti{\'e}rrez and P.~Le~Callet, ``{Salient360!}: Visual attention modeling
  for 360{\dg} content,'' in \emph{{IEEE} International Conference on
  Multimedia and Expo 2018 Grand Challenge}.\hskip 1em plus 0.5em minus
  0.4em\relax IEEE, 2018.

\bibitem{de2017look}
A.~De~Abreu, C.~Ozcinar, and A.~Smolic, ``Look around you: Saliency maps for
  omnidirectional images in {VR} applications,'' in \emph{International
  Conference on Quality of Multimedia Experience}.\hskip 1em plus 0.5em minus
  0.4em\relax IEEE, 2017, pp. 1--6.

\bibitem{bao2016shooting}
Y.~Bao, H.~Wu, T.~Zhang, A.~A. Ramli, and X.~Liu, ``Shooting a moving target:
  Motion-prediction-based transmission for 360-degree videos,'' in \emph{{IEEE}
  International Conference on Big Data}.\hskip 1em plus 0.5em minus 0.4em\relax
  IEEE, 2016, pp. 1161--1170.

\bibitem{ozcinar2018visual}
C.~Ozcinar and A.~Smolic, ``Visual attention in omnidirectional video for
  virtual reality applications,'' in \emph{International Conference on Quality
  of Multimedia Experience}.\hskip 1em plus 0.5em minus 0.4em\relax IEEE, 2018,
  pp. 1--6.

\bibitem{corbillon2017360}
X.~Corbillon, F.~De~Simone, and G.~Simon, ``360-degree video head movement
  dataset,'' in \emph{{ACM} Multimedia Systems Conference}.\hskip 1em plus
  0.5em minus 0.4em\relax ACM, 2017, pp. 199--204.

\bibitem{lo2017360}
W.-C. Lo, C.-L. Fan, J.~Lee, C.-Y. Huang, K.-T. Chen, and C.-H. Hsu, ``360
  video viewing dataset in head-mounted virtual reality,'' in \emph{{ACM}
  Multimedia Systems Conference}.\hskip 1em plus 0.5em minus 0.4em\relax ACM,
  2017, pp. 211--216.

\bibitem{wu2017dataset}
C.~Wu, Z.~Tan, Z.~Wang, and S.~Yang, ``A dataset for exploring user behaviors
  in {VR} spherical video streaming,'' in \emph{{ACM} Multimedia Systems
  Conference}.\hskip 1em plus 0.5em minus 0.4em\relax ACM, 2017, pp. 193--198.

\bibitem{fremerey2018avtrack360}
S.~Fremerey, A.~Singla, K.~Meseberg, and A.~Raake, ``{AVtrack360}: an open
  dataset and software recording people's head rotations watching 360{\dg}
  videos on an {HMD},'' in \emph{{ACM} Multimedia Systems Conference}.\hskip
  1em plus 0.5em minus 0.4em\relax ACM, 2018, pp. 403--408.

\bibitem{cheng2018cube}
H.-T. Cheng, C.-H. Chao, J.-D. Dong, H.-K. Wen, T.-L. Liu, and M.~Sun, ``Cube
  padding for weakly-supervised saliency prediction in 360{\dg} videos,'' in
  \emph{{IEEE} Conference on Computer Vision and Pattern Recognition}.\hskip
  1em plus 0.5em minus 0.4em\relax IEEE, 2018, pp. 1420--1429.

\bibitem{sitzmann2018saliency}
V.~Sitzmann, A.~Serrano, A.~Pavel, M.~Agrawala, D.~Gutierrez, B.~Masia, and
  G.~Wetzstein, ``Saliency in {VR}: How do people explore virtual
  environments?'' \emph{{IEEE} Transactions on Visualization and Computer
  Graphics}, vol.~24, no.~4, pp. 1633--1642, 2018.

\bibitem{rai2017dataset}
Y.~Rai, J.~Guti{\'e}rrez, and P.~Le~Callet, ``A dataset of head and eye
  movements for 360 degree images,'' in \emph{{ACM} Multimedia Systems
  Conference}.\hskip 1em plus 0.5em minus 0.4em\relax ACM, 2017, pp. 205--210.

\bibitem{david2018dataset}
E.~J. David, J.~Guti{\'e}rrez, A.~Coutrot, M.~P. Da~Silva, and P.~L. Callet,
  ``A dataset of head and eye movements for 360{\dg} videos,'' in \emph{{ACM}
  Multimedia Systems Conference}.\hskip 1em plus 0.5em minus 0.4em\relax ACM,
  2018, pp. 432--437.

\bibitem{xu2018predicting}
M.~Xu, Y.~Song, J.~Wang, M.~Qiao, L.~Huo, and Z.~Wang, ``Predicting head
  movement in panoramic video: A deep reinforcement learning approach,''
  \emph{{IEEE} Transactions on Pattern Analysis and Machine Intelligence},
  2018.

\bibitem{zhang2018saliency}
Z.~Zhang, Y.~Xu, J.~Yu, and S.~Gao, ``Saliency detection in 360{\dg} videos,''
  in \emph{European Conference on Computer Vision}.\hskip 1em plus 0.5em minus
  0.4em\relax Springer, 2018, pp. 504--520.

\bibitem{hu2017deep}
H.-N. Hu, Y.-C. Lin, M.-Y. Liu, H.-T. Cheng, Y.-J. Chang, and M.~Sun, ``Deep
  360 pilot: Learning a deep agent for piloting through 360 sports videos,'' in
  \emph{{IEEE} Conference on Computer Vision and Pattern Recognition}.\hskip
  1em plus 0.5em minus 0.4em\relax IEEE, 2017, pp. 1396--1405.

\bibitem{xu2018gaze}
Y.~Xu, Y.~Dong, J.~Wu, Z.~Sun, Z.~Shi, J.~Yu, and S.~Gao, ``Gaze prediction in
  dynamic 360{\dg} immersive videos,'' in \emph{{IEEE} Conference on Computer
  Vision and Pattern Recognition}.\hskip 1em plus 0.5em minus 0.4em\relax IEEE,
  2018, pp. 5333--5342.

\bibitem{li2018bridge}
C.~Li, M.~Xu, X.~Du, and Z.~Wang, ``Bridge the gap between {VQA} and human
  behavior on omnidirectional video: A large-scale dataset and a deep learning
  model,'' in \emph{{ACM} International Conference on Multimedia}.\hskip 1em
  plus 0.5em minus 0.4em\relax ACM, 2018, pp. 932--940.

\bibitem{startsev2018360}
M.~Startsev and M.~Dorr, ``360-aware saliency estimation with conventional
  image saliency predictors,'' \emph{Signal Processing: Image Communication},
  vol.~69, pp. 43--52, 2018.

\bibitem{lebreton2018gbvs360}
P.~Lebreton and A.~Raake, ``{GBVS360}, {BMS360}, {ProSal}: Extending existing
  saliency prediction models from {2D} to omnidirectional images,''
  \emph{Signal Processing: Image Communication}, vol.~69, pp. 69--78, 2018.

\bibitem{luz2017saliency}
G.~Luz, J.~Ascenso, C.~Brites, and F.~Pereira, ``Saliency-driven
  omnidirectional imaging adaptive coding: Modeling and assessment,'' in
  \emph{{IEEE} International Workshop on Multimedia Signal Processing}.\hskip
  1em plus 0.5em minus 0.4em\relax IEEE, 2017, pp. 1--6.

\bibitem{battisti2018feature}
F.~Battisti, S.~Baldoni, M.~Brizzi, and M.~Carli, ``A feature-based approach
  for saliency estimation of omni-directional images,'' \emph{Signal
  Processing: Image Communication}, vol.~69, pp. 53--59, 2018.

\bibitem{zhu2018prediction}
Y.~Zhu, G.~Zhai, and X.~Min, ``The prediction of head and eye movement for 360
  degree images,'' \emph{Signal Processing: Image Communication}, vol.~69, pp.
  15--25, 2018.

\bibitem{fang2018novel}
Y.~Fang, X.~Zhang, and N.~Imamoglu, ``A novel superpixel-based saliency
  detection model for 360-degree images,'' \emph{Signal Processing: Image
  Communication}, vol.~69, pp. 1--7, 2018.

\bibitem{ling2018saliency}
J.~Ling, K.~Zhang, Y.~Zhang, D.~Yang, and Z.~Chen, ``A saliency prediction
  model on 360 degree images using color dictionary based sparse
  representation,'' \emph{Signal Processing: Image Communication}, vol.~69, pp.
  60--68, 2018.

\bibitem{assens2018scanpath}
M.~Assens, X.~Giro-i Nieto, K.~McGuinness, and N.~E. O’Connor, ``Scanpath and
  saliency prediction on 360 degree images,'' \emph{Signal Processing: Image
  Communication}, vol.~69, pp. 8--14, 2018.

\bibitem{monroy2018salnet360}
R.~Monroy, S.~Lutz, T.~Chalasani, and A.~Smolic, ``{SalNet360}: Saliency maps
  for omni-directional images with {CNN},'' \emph{Signal Processing: Image
  Communication}, vol.~69, pp. 26--34, 2018.

\bibitem{chao2018salgan360}
F.-Y. Chao, L.~Zhang, W.~Hamidouche, and O.~Deforges, ``{SalGAN360}: Visual
  saliency prediction on 360 degree images with generative adversarial
  networks,'' in \emph{{IEEE} International Conference on Multimedia and Expo
  Workshops}.\hskip 1em plus 0.5em minus 0.4em\relax IEEE, 2018, pp. 01--04.

\bibitem{suzuki2018saliency}
T.~Suzuki and T.~Yamanaka, ``Saliency map estimation for omni-directional image
  considering prior distributions,'' in \emph{{IEEE} International Conference
  on Systems, Man, and Cybernetics}.\hskip 1em plus 0.5em minus 0.4em\relax
  IEEE, 2018, pp. 2079--2084.

\bibitem{lebreton2018v}
P.~Lebreton, S.~Fremerey, and A.~Raake, ``{V-BMS360}: A video extention to the
  {BMS360} image saliency model,'' in \emph{{IEEE} International Conference on
  Multimedia and Expo Workshops}.\hskip 1em plus 0.5em minus 0.4em\relax IEEE,
  2018, pp. 1--4.

\bibitem{xu2018assessing}
M.~Xu, C.~Li, Z.~Chen, Z.~Wang, and Z.~Guan, ``Assessing visual quality of
  omnidirectional videos,'' \emph{{IEEE} Transactions on Circuits and Systems
  for Video Technology}, 2018.

\bibitem{rai2017saliency}
Y.~Rai, P.~Le~Callet, and P.~Guillotel, ``Which saliency weighting for omni
  directional image quality assessment?'' in \emph{International Conference on
  Quality of Multimedia Experience}.\hskip 1em plus 0.5em minus 0.4em\relax
  IEEE, 2017, pp. 1--6.

\bibitem{nguyen2018your}
A.~Nguyen, Z.~Yan, and K.~Nahrstedt, ``Your attention is unique: Detecting
  360-degree video saliency in head-mounted display for head movement
  prediction,'' in \emph{{ACM} International Conference on Multimedia}.\hskip
  1em plus 0.5em minus 0.4em\relax ACM, 2018, pp. 1190--1198.

\bibitem{aladagli2017predicting}
A.~D. Aladagli, E.~Ekmekcioglu, D.~Jarnikov, and A.~Kondoz, ``Predicting head
  trajectories in 360{\dg} virtual reality videos,'' in \emph{International
  Conference on {3D} Immersion}.\hskip 1em plus 0.5em minus 0.4em\relax IEEE,
  2017, pp. 1--6.

\bibitem{assens2017saltinet}
M.~Assens, X.~Giro-i Nieto, K.~McGuinness, and N.~E. O’Connor, ``{SaltiNet}:
  Scan-path prediction on 360 degree images using saliency volumes,'' in
  \emph{{IEEE} International Conference on Computer Vision Workshops}.\hskip
  1em plus 0.5em minus 0.4em\relax IEEE, 2017, pp. 2331--2338.

\bibitem{fan2017fixation}
C.-L. Fan, J.~Lee, W.-C. Lo, C.-Y. Huang, K.-T. Chen, and C.-H. Hsu, ``Fixation
  prediction for 360 video streaming in head-mounted virtual reality,'' in
  \emph{{ACM} {SIGMM} Workshop on Network and Operating Systems Support for
  Digital Audio and Video}.\hskip 1em plus 0.5em minus 0.4em\relax ACM, 2017,
  pp. 67--72.

\bibitem{bogdanova2008spherical}
I.~Bogdanova, A.~Bur, and H.~H{\"u}gli, ``The spherical approach to
  omnidirectional visual attention,'' in \emph{European Signal Processing
  Conference}.\hskip 1em plus 0.5em minus 0.4em\relax IEEE, 2008, pp. 1--5.

\bibitem{bogdanova2008visual}
I.~Bogdanova, A.~Bur, and H.~Hugli, ``Visual attention on the sphere,''
  \emph{{IEEE} Transactions on Image Processing}, vol.~17, no.~11, pp.
  2000--2014, 2008.

\bibitem{cohen2018spherical}
\BIBentryALTinterwordspacing
T.~S. Cohen, M.~Geiger, J.~Köhler, and M.~Welling, ``Spherical {CNN}s,'' in
  \emph{International Conference on Learning Representations}, 2018. [Online].
  Available: \url{https://openreview.net/forum?id=Hkbd5xZRb}
\BIBentrySTDinterwordspacing

\bibitem{koch1987shifts}
C.~Koch and S.~Ullman, ``Shifts in selective visual attention: towards the
  underlying neural circuitry,'' in \emph{Matters of intelligence}.\hskip 1em
  plus 0.5em minus 0.4em\relax Springer, 1987, pp. 115--141.

\bibitem{bogdanova2010dynamic}
I.~Bogdanova, A.~Bur, H.~H{\"u}gli, and P.-A. Farine, ``Dynamic visual
  attention on the sphere,'' \emph{Computer Vision and Image Understanding},
  vol. 114, no.~1, pp. 100--110, 2010.

\bibitem{borji2013state}
A.~Borji and L.~Itti, ``State-of-the-art in visual attention modeling,''
  \emph{{IEEE} Transactions on Pattern Analysis and Machine Intelligence},
  vol.~35, no.~1, pp. 185--207, 2013.

\bibitem{maugey2017saliency}
T.~Maugey, O.~Le~Meur, and Z.~Liu, ``Saliency-based navigation in
  omnidirectional image,'' in \emph{{IEEE} International Workshop on Multimedia
  Signal Processing}.\hskip 1em plus 0.5em minus 0.4em\relax IEEE, 2017, pp.
  1--6.

\bibitem{huang2015salicon}
X.~Huang, C.~Shen, X.~Boix, and Q.~Zhao, ``{SALICON}: Reducing the semantic gap
  in saliency prediction by adapting deep neural networks,'' in \emph{{IEEE}
  International Conference on Computer Vision}.\hskip 1em plus 0.5em minus
  0.4em\relax IEEE, 2015, pp. 262--270.

\bibitem{zhang2016exploiting}
J.~Zhang and S.~Sclaroff, ``Exploiting surroundedness for saliency detection: a
  boolean map approach,'' \emph{{IEEE} Transactions on Pattern Analysis and
  Machine Intelligence}, vol.~38, no.~5, pp. 889--902, 2016.

\bibitem{snyder1987map}
J.~P. Snyder, \emph{Map projections--A working manual}.\hskip 1em plus 0.5em
  minus 0.4em\relax US Government Printing Office, 1987, vol. 1395.

\bibitem{harel2006graph}
J.~Harel, C.~Koch, and P.~Perona, ``Graph-based visual saliency,'' in
  \emph{Advances in Neural Information Processing Systems}.\hskip 1em plus
  0.5em minus 0.4em\relax MIT Press, 2006, pp. 545--552.

\bibitem{cornia2016deep}
M.~Cornia, L.~Baraldi, G.~Serra, and R.~Cucchiara, ``A deep multi-level network
  for saliency prediction,'' in \emph{International Conference on Pattern
  Recognition}.\hskip 1em plus 0.5em minus 0.4em\relax IEEE, 2016, pp.
  3488--3493.

\bibitem{thakur2011face}
S.~Thakur, S.~Paul, A.~Mondal, S.~Das, and A.~Abraham, ``Face detection using
  skin tone segmentation,'' in \emph{World Congress on Information and
  Communication Technologies}.\hskip 1em plus 0.5em minus 0.4em\relax IEEE,
  2011, pp. 53--60.

\bibitem{viola2004robust}
P.~Viola and M.~J. Jones, ``Robust real-time face detection,''
  \emph{International Journal of Computer Vision}, vol.~57, no.~2, pp.
  137--154, 2004.

\bibitem{achanta2012slic}
R.~Achanta, A.~Shaji, K.~Smith, A.~Lucchi, P.~Fua, and S.~S{\"u}sstrunk,
  ``{SLIC} superpixels compared to state-of-the-art superpixel methods,''
  \emph{{IEEE} Transactions on Pattern Analysis and Machine Intelligence},
  vol.~34, no.~11, pp. 2274--2282, 2012.

\bibitem{judd2009learning}
T.~Judd, K.~Ehinger, F.~Durand, and A.~Torralba, ``Learning to predict where
  humans look,'' in \emph{{IEEE} International Conference on Computer
  Vision}.\hskip 1em plus 0.5em minus 0.4em\relax IEEE, 2009, pp. 2106--2113.

\bibitem{goodfellow2016deep}
I.~Goodfellow, Y.~Bengio, and A.~Courville, \emph{Deep learning}.\hskip 1em
  plus 0.5em minus 0.4em\relax MIT press, 2016.

\bibitem{borji2018saliency}
A.~Borji, ``Saliency prediction in the deep learning era: An empirical
  investigation,'' \emph{arXiv preprint arXiv:1810.03716}, 2018.

\bibitem{pan2016shallow}
J.~Pan, E.~Sayrol, X.~Giro-I-Nieto, K.~McGuinness, and N.~E. O’Connor,
  ``Shallow and deep convolutional networks for saliency prediction,'' in
  \emph{{IEEE} Conference on Computer Vision and Pattern Recognition}.\hskip
  1em plus 0.5em minus 0.4em\relax IEEE, 2016, pp. 598--606.

\bibitem{pan2017salgan}
J.~Pan, C.~C. Ferrer, K.~McGuinness, N.~E. O'Connor, J.~Torres, E.~Sayrol, and
  X.~Giro-i Nieto, ``Salgan: Visual saliency prediction with generative
  adversarial networks,'' \emph{arXiv preprint arXiv:1701.01081}, 2017.

\bibitem{revesz2005random}
P.~R{\'e}v{\'e}sz, \emph{Random walk in random and non-random
  environments}.\hskip 1em plus 0.5em minus 0.4em\relax World Scientific, 2005.

\bibitem{zakharchenko2016quality}
V.~Zakharchenko, K.~P. Choi, and J.~H. Park, ``Quality metric for spherical
  panoramic video,'' in \emph{Optics and Photonics for Information Processing},
  vol. 9970.\hskip 1em plus 0.5em minus 0.4em\relax International Society for
  Optics and Photonics, 2016, p. 99700C.

\bibitem{boyce2017jvet}
J.~Boyce, E.~Alshina, A.~Abbas, and Y.~Ye, ``{JVET} common test conditions and
  evaluation procedures for 360{\dg} video,'' \emph{Joint Video Exploration
  Team of {ITU-T SG}}, vol.~16, 2017.

\bibitem{hanhart2018360}
P.~Hanhart, Y.~He, Y.~Ye, J.~Boyce, Z.~Deng, and L.~Xu, ``360-degree video
  quality evaluation,'' in \emph{Picture Coding Symposium}.\hskip 1em plus
  0.5em minus 0.4em\relax IEEE, 2018, pp. 328--332.

\bibitem{huang2018modeling}
M.~Huang, Q.~Shen, Z.~Ma, A.~C. Bovik, P.~Gupta, R.~Zhou, and X.~Cao,
  ``Modeling the perceptual quality of immersive images rendered on head
  mounted displays: Resolution and compression,'' \emph{{IEEE} Transactions on
  Image Processing}, vol.~27, no.~12, pp. 6039--6050, 2018.

\bibitem{duan2017ivqad}
H.~Duan, G.~Zhai, X.~Yang, D.~Li, and W.~Zhu, ``Ivqad 2017: An immersive video
  quality assessment database,'' in \emph{International Conference on Systems,
  Signals and Image Processing}.\hskip 1em plus 0.5em minus 0.4em\relax IEEE,
  2017, pp. 1--5.

\bibitem{zhang2017subjective}
B.~Zhang, J.~Zhao, S.~Yang, Y.~Zhang, J.~Wang, and Z.~Fei, ``Subjective and
  objective quality assessment of panoramic videos in virtual reality
  environments,'' in \emph{{IEEE} International Conference on Multimedia and
  Expo Workshops}.\hskip 1em plus 0.5em minus 0.4em\relax IEEE, 2017, pp.
  163--168.

\bibitem{lopes2018subjective}
F.~Lopes, J.~Ascenso, A.~Rodrigues, and M.~P. Queluz, ``Subjective and
  objective quality assessment of omnidirectional video,'' in
  \emph{Applications of Digital Image Processing}, vol. 10752.\hskip 1em plus
  0.5em minus 0.4em\relax International Society for Optics and Photonics, 2018,
  p. 107520P.

\bibitem{singla2017comparison}
A.~Singla, S.~Fremerey, W.~Robitza, P.~Lebreton, and A.~Raake, ``Comparison of
  subjective quality evaluation for {HEVC} encoded omnidirectional videos at
  different bit-rates for {UHD} and {FHD} resolution,'' in \emph{Thematic
  Workshops of {ACM} Multimedia}.\hskip 1em plus 0.5em minus 0.4em\relax ACM,
  2017, pp. 511--519.

\bibitem{sun2017cviqd}
W.~Sun, K.~Gu, G.~Zhai, S.~Ma, W.~Lin, and P.~Le~Calle, ``{CVIQD}: Subjective
  quality evaluation of compressed virtual reality images,'' in \emph{{IEEE}
  International Conference on Image Processing}.\hskip 1em plus 0.5em minus
  0.4em\relax IEEE, 2017, pp. 3450--3454.

\bibitem{sun2018large}
W.~Sun, K.~Gu, S.~Ma, W.~Zhu, N.~Liu, and G.~Zhai, ``A large-scale compressed
  360-degree spherical image database: From subjective quality evaluation to
  objective model comparison,'' in \emph{{IEEE} International Workshop on
  Multimedia Signal Processing}.\hskip 1em plus 0.5em minus 0.4em\relax IEEE,
  2018, pp. 1--6.

\bibitem{duan2018perceptual}
H.~Duan, G.~Zhai, X.~Min, Y.~Zhu, Y.~Fang, and X.~Yang, ``Perceptual quality
  assessment of omnidirectional images,'' in \emph{{IEEE} International
  Symposium on Circuits and Systems}.\hskip 1em plus 0.5em minus 0.4em\relax
  IEEE, 2018, pp. 1--5.

\bibitem{upenik2017performance}
E.~Upenik, M.~Rerabek, and T.~Ebrahimi, ``On the performance of objective
  metrics for omnidirectional visual content,'' in \emph{International
  Conference on Quality of Multimedia Experience}.\hskip 1em plus 0.5em minus
  0.4em\relax IEEE, 2017, pp. 1--6.

\bibitem{chen2018towards}
Z.~Chen and Y.~Zhang, ``Towards subjective quality assessment for panoramic
  video,'' \emph{{IS\&T} International Symposium on Electronic Imaging}, vol.
  2018, no.~14, pp. 1--5, 2018.

\bibitem{zhang2018subjective}
Y.~Zhang, Y.~Wang, F.~Liu, Z.~Liu, Y.~Li, D.~Yang, and Z.~Chen, ``Subjective
  panoramic video quality assessment database for coding applications,''
  \emph{{IEEE} Transactions on Broadcasting}, vol.~64, no.~2, pp. 461--473,
  2018.

\bibitem{tran2018study}
H.~T. Tran, C.~T. Pham, N.~P. Ngoc, A.~T. Pham, and T.~C. Thang, ``A study on
  quality metrics for 360 video communications,'' \emph{{IEICE} Transactions on
  Information and Systems}, vol. 101, no.~1, pp. 28--36, 2018.

\bibitem{itut2006mean}
{ITU-T}, ``Mean opinion score ({MOS}) terminology,'' \emph{Recommendation
  {ITU-T P.800.1}}, 2006.

\bibitem{yu2015framework}
M.~Yu, H.~Lakshman, and B.~Girod, ``A framework to evaluate omnidirectional
  video coding schemes,'' in \emph{{IEEE} International Symposium on Mixed and
  Augmented Reality}.\hskip 1em plus 0.5em minus 0.4em\relax IEEE, 2015, pp.
  31--36.

\bibitem{zhang2014vsi}
L.~Zhang, Y.~Shen, and H.~Li, ``{VSI}: A visual saliency-induced index for
  perceptual image quality assessment,'' \emph{{IEEE} Transactions on Image
  Processing}, vol.~23, no.~10, pp. 4270--4281, 2014.

\bibitem{zhang2017study}
W.~Zhang and H.~Liu, ``Study of saliency in objective video quality
  assessment,'' \emph{{IEEE} Transactions on Image Processing}, vol.~26, no.~3,
  pp. 1275--1288, 2017.

\bibitem{wang2004image}
Z.~Wang, A.~C. Bovik, H.~R. Sheikh, E.~P. Simoncelli \emph{et~al.}, ``Image
  quality assessment: from error visibility to structural similarity,''
  \emph{{IEEE} Transactions on Image Processing}, vol.~13, no.~4, pp. 600--612,
  2004.

\bibitem{zakharchenko2017omnidirectional}
V.~Zakharchenko, K.~P. Choi, E.~Alshina, and J.~H. Park, ``Omnidirectional
  video quality metrics and evaluation process,'' in \emph{Data Compression
  Conference}.\hskip 1em plus 0.5em minus 0.4em\relax IEEE, 2017, pp. 472--472.

\bibitem{youvalari2016analysis}
R.~G. Youvalari, A.~Aminlou, and M.~M. Hannuksela, ``Analysis of regional
  down-sampling methods for coding of omnidirectional video,'' in \emph{Picture
  Coding Symposium}.\hskip 1em plus 0.5em minus 0.4em\relax IEEE, 2016, pp.
  1--5.

\bibitem{sun2017weighted}
Y.~Sun, A.~Lu, and L.~Yu, ``Weighted-to-spherically-uniform quality evaluation
  for omnidirectional video,'' \emph{{IEEE} Signal Processing Letters},
  vol.~24, no.~9, pp. 1408--1412, 2017.

\bibitem{xiu2017evaluation}
X.~Xiu, Y.~He, Y.~Ye, and B.~Vishwanath, ``An evaluation framework for
  360-degree video compression,'' in \emph{{IEEE} Visual Communications and
  Image Processing}.\hskip 1em plus 0.5em minus 0.4em\relax IEEE, 2017, pp.
  1--4.

\bibitem{wang2003multiscale}
Z.~Wang, E.~P. Simoncelli, and A.~C. Bovik, ``Multi-scale structural similarity
  for image quality assessment,'' in \emph{Asilomar Conference on Signals,
  Systems and Computers}, vol.~2.\hskip 1em plus 0.5em minus 0.4em\relax IEEE,
  2003, pp. 1398--1402.

\bibitem{zhou2018weighted}
Y.~Zhou, M.~Yu, H.~Ma, H.~Shao, and G.~Jiang, ``Weighted-to-spherically-uniform
  ssim objective quality evaluation for panoramic video,'' in \emph{{IEEE}
  International Conference on Signal Processing}.\hskip 1em plus 0.5em minus
  0.4em\relax IEEE, 2018, pp. 54--57.

\bibitem{chen2018spherical}
S.~Chen, Y.~Zhang, Y.~Li, Z.~Chen, and Z.~Wang, ``Spherical structural
  similarity index for objective omnidirectional video quality assessment,'' in
  \emph{{IEEE} International Conference on Multimedia and Expo}.\hskip 1em plus
  0.5em minus 0.4em\relax IEEE, 2018, pp. 1--6.

\bibitem{ma2008image}
Q.~Ma and L.~Zhang, ``Image quality assessment with visual attention,'' in
  \emph{International Conference on Pattern Recognition}.\hskip 1em plus 0.5em
  minus 0.4em\relax IEEE, 2008, pp. 1--4.

\bibitem{ma2008saliency}
------, ``Saliency-based image quality assessment criterion,'' in
  \emph{International Conference on Intelligent Computing}.\hskip 1em plus
  0.5em minus 0.4em\relax Springer, 2008, pp. 1124--1133.

\bibitem{ozcinar2019visual}
C.~Ozcinar, J.~Cabrera, and A.~Smolic, ``Visual attention-aware omnidirectional
  video streaming using optimal tiles for virtual reality,'' \emph{{IEEE}
  Journal on Emerging and Selected Topics in Circuits and Systems}, vol.~9,
  no.~1, pp. 217--230, 2019.

\bibitem{yang2017objective}
S.~Yang, J.~Zhao, T.~Jiang, J.~W.~T. Rahim, B.~Zhang, Z.~Xu, and Z.~Fei, ``An
  objective assessment method based on multi-level factors for panoramic
  videos,'' in \emph{{IEEE} Visual Communications and Image Processing}.\hskip
  1em plus 0.5em minus 0.4em\relax IEEE, 2017, pp. 1--4.

\bibitem{murray2011saliency}
N.~Murray, M.~Vanrell, X.~Otazu, and C.~A. Parraga, ``Saliency estimation using
  a non-parametric low-level vision model,'' in \emph{{IEEE} Conference on
  Computer Vision and Pattern Recognition}.\hskip 1em plus 0.5em minus
  0.4em\relax IEEE, 2011, pp. 433--440.

\bibitem{yang2013saliency}
C.~Yang, L.~Zhang, H.~Lu, X.~Ruan, and M.-H. Yang, ``Saliency detection via
  graph-based manifold ranking,'' in \emph{{IEEE} Conference on Computer Vision
  and Pattern Recognition}.\hskip 1em plus 0.5em minus 0.4em\relax IEEE, 2013,
  pp. 3166--3173.

\bibitem{shelhamer2017fully}
E.~Shelhamer, J.~Long, and T.~Darrell, ``Fully convolutional networks for
  semantic segmentation,'' \emph{{IEEE} Transactions on Pattern Analysis and
  Machine Intelligence}, vol.~39, no.~4, pp. 640--651, 2017.

\bibitem{ou2011perceptual}
Y.-F. Ou, Z.~Ma, T.~Liu, and Y.~Wang, ``Perceptual quality assessment of video
  considering both frame rate and quantization artifacts,'' \emph{{IEEE}
  Transactions on Circuits and Systems for Video Technology}, vol.~21, no.~3,
  pp. 286--298, 2011.

\bibitem{lim2018vr}
H.-T. Lim, H.~G. Kim, and Y.~M. Ra, ``Vr iqa net: Deep virtual reality image
  quality assessment using adversarial learning,'' in \emph{{IEEE}
  International Conference on Acoustics, Speech and Signal Processing}.\hskip
  1em plus 0.5em minus 0.4em\relax IEEE, 2018, pp. 6737--6741.

\bibitem{kim2019deep}
H.~G. Kim, H.-t. Lim, and Y.~M. Ro, ``Deep virtual reality image quality
  assessment with human perception guider for omnidirectional image,''
  \emph{{IEEE} Transactions on Circuits and Systems for Video Technology},
  2019.

\bibitem{li2019viewport}
C.~Li, M.~Xu, L.~Jiang, S.~Zhang, and X.~Tao, ``Viewport proposal {CNN} for
  360{\dg} video quality assessment,'' in \emph{{IEEE} Conference on Computer
  Vision and Pattern Recognition}.\hskip 1em plus 0.5em minus 0.4em\relax IEEE,
  2019.

\bibitem{bosse2018deep}
S.~Bosse, D.~Maniry, K.-R. M{\"u}ller, T.~Wiegand, and W.~Samek, ``Deep neural
  networks for no-reference and full-reference image quality assessment,''
  \emph{{IEEE} Transactions on Image Processing}, vol.~27, no.~1, pp. 206--219,
  2018.

\bibitem{pan2018blind}
D.~Pan, P.~Shi, M.~Hou, Z.~Ying, S.~Fu, and Y.~Zhang, ``Blind predicting
  similar quality map for image quality assessment,'' in \emph{{IEEE}
  Conference on Computer Vision and Pattern Recognition}.\hskip 1em plus 0.5em
  minus 0.4em\relax IEEE, 2018, pp. 6373--6382.

\bibitem{lin2018hallucinated}
K.-Y. Lin and G.~Wang, ``Hallucinated-{IQA}: No-reference image quality
  assessment via adversarial learning,'' in \emph{{IEEE} Conference on Computer
  Vision and Pattern Recognition}.\hskip 1em plus 0.5em minus 0.4em\relax IEEE,
  2018, pp. 732--741.

\bibitem{kim2018deep}
J.~Kim, A.-D. Nguyen, and S.~Lee, ``Deep {CNN}-based blind image quality
  predictor,'' \emph{{IEEE} Transactions on Neural Networks and Learning
  Systems}, vol.~30, no.~1, pp. 11--24, 2019.

\bibitem{huang2017densely}
G.~Huang, Z.~Liu, L.~van~der Maaten, and K.~Q. Weinberger, ``Densely connected
  convolutional networks,'' in \emph{{IEEE} Conference on Computer Vision and
  Pattern Recognition}.\hskip 1em plus 0.5em minus 0.4em\relax IEEE, 2017, pp.
  2261--2269.

\bibitem{wang2017new}
Y.~Wang, L.~Li, D.~Liu, F.~Wu, and W.~Gao, ``A new motion model for panoramic
  video coding,'' in \emph{{IEEE} International Conference on Image
  Processing}.\hskip 1em plus 0.5em minus 0.4em\relax IEEE, 2017, pp.
  1407--1411.

\bibitem{ghaznavi2018geometry}
R.~Ghaznavi-Youvalari and A.~Aminlou, ``Geometry-based motion vector scaling
  for omnidirectional video coding,'' in \emph{{IEEE} International Symposium
  on Multimedia}.\hskip 1em plus 0.5em minus 0.4em\relax IEEE, 2018, pp.
  127--130.

\bibitem{he2017motion}
Y.~He, Y.~Ye, P.~Hanhart, and X.~Xiu, ``Motion compensated prediction with
  geometry padding for 360 video coding,'' in \emph{{IEEE} Visual
  Communications and Image Processing}.\hskip 1em plus 0.5em minus 0.4em\relax
  IEEE, 2017, pp. 1--4.

\bibitem{li2019advanced}
L.~Li, Z.~Li, X.~Ma, H.~Yang, and H.~Li, ``Advanced spherical motion model and
  local padding for 360{\dg} video compression,'' \emph{{IEEE} Transactions on
  Image Processing}, vol.~28, no.~5, pp. 2342--2356, 2019.

\bibitem{tosic2005multiresolution}
I.~Tosic, I.~Bogdanova, P.~Frossard, and P.~Vandergheynst, ``Multiresolution
  motion estimation for omnidirectional images,'' in \emph{European Signal
  Processing Conference}.\hskip 1em plus 0.5em minus 0.4em\relax IEEE, 2005,
  pp. 1--4.

\bibitem{li2017projection}
L.~Li, Z.~Li, M.~Budagavi, and H.~Li, ``Projection based advanced motion model
  for cubic mapping for 360-degree video,'' in \emph{{IEEE} International
  Conference on Image Processing}.\hskip 1em plus 0.5em minus 0.4em\relax IEEE,
  2017, pp. 1427--1431.

\bibitem{vishwanath2017rotational}
B.~Vishwanath, T.~Nanjundaswamy, and K.~Rose, ``Rotational motion model for
  temporal prediction in 360 video coding,'' in \emph{{IEEE} International
  Workshop on Multimedia Signal Processing}.\hskip 1em plus 0.5em minus
  0.4em\relax IEEE, 2017, pp. 1--6.

\bibitem{de2017deformable}
F.~De~Simone, P.~Frossard, N.~Birkbeck, and B.~Adsumilli, ``Deformable
  block-based motion estimation in omnidirectional image sequences,'' in
  \emph{{IEEE} International Workshop on Multimedia Signal Processing}.\hskip
  1em plus 0.5em minus 0.4em\relax IEEE, 2017, pp. 1--6.

\bibitem{wang2019spherical}
Y.~Wang, D.~Liu, S.~Ma, F.~Wu, and W.~Gao, ``Spherical coordinates
  transform-based motion model for panoramic video coding,'' \emph{{IEEE}
  Journal on Emerging and Selected Topics in Circuits and Systems}, vol.~9,
  no.~1, pp. 98--109, 2019.

\bibitem{zheng2007adaptive}
J.~Zheng, Y.~Shen, Y.~Zhang, and G.~Ni, ``Adaptive selection of motion models
  for panoramic video coding,'' in \emph{{IEEE} International Conference on
  Multimedia and Expo}.\hskip 1em plus 0.5em minus 0.4em\relax IEEE, 2007, pp.
  1319--1322.

\bibitem{sauer2017improved}
J.~Sauer, J.~Schneider, and M.~Wien, ``Improved motion compensation for
  360{\dg} video projected to polytopes,'' in \emph{{IEEE} International
  Conference on Multimedia and Expo}.\hskip 1em plus 0.5em minus 0.4em\relax
  IEEE, 2017, pp. 61--66.

\bibitem{sullivan2012overview}
G.~J. Sullivan, J.-R. Ohm, W.-J. Han, and T.~Wiegand, ``Overview of the high
  efficiency video coding ({HEVC}) standard,'' \emph{{IEEE} Transactions on
  Circuits and Systems for Video Technology}, vol.~22, no.~12, pp. 1649--1668,
  2012.

\bibitem{li2018reference}
N.~Li, S.~Wan, and F.~Yang, ``Reference samples padding for intra-frame coding
  of omnidirectional video,'' in \emph{Asia-Pacific Signal and Information
  Processing Association Annual Summit and Conference}.\hskip 1em plus 0.5em
  minus 0.4em\relax IEEE, 2018, pp. 1987--1990.

\bibitem{budagavi2015360}
M.~Budagavi, J.~Furton, G.~Jin, A.~Saxena, J.~Wilkinson, and A.~Dickerson,
  ``360 degrees video coding using region adaptive smoothing,'' in \emph{{IEEE}
  International Conference on Image Processing}.\hskip 1em plus 0.5em minus
  0.4em\relax IEEE, 2015, pp. 750--754.

\bibitem{lee2017omnidirectional}
S.-H. Lee, S.-T. Kim, E.~Yip, B.-D. Choi, J.~Song, and S.-J. Ko,
  ``Omnidirectional video coding using latitude adaptive down-sampling and
  pixel rearrangement,'' \emph{Electronics Letters}, vol.~53, no.~10, pp.
  655--657, 2017.

\bibitem{xiu2018adaptive}
X.~Xiu, Y.~He, and Y.~Ye, ``An adaptive quantization method for 360-degree
  video coding,'' in \emph{Applications of Digital Image Processing}, vol.
  10752.\hskip 1em plus 0.5em minus 0.4em\relax International Society for
  Optics and Photonics, 2018, p. 107520X.

\bibitem{liu2017novel}
Y.~Liu, M.~Xu, C.~Li, S.~Li, and Z.~Wang, ``A novel rate control scheme for
  panoramic video coding,'' in \emph{{IEEE} International Conference on
  Multimedia and Expo}.\hskip 1em plus 0.5em minus 0.4em\relax IEEE, 2017, pp.
  691--696.

\bibitem{tang2017optimized}
M.~Tang, Y.~Zhang, J.~Wen, and S.~Yang, ``Optimized video coding for
  omnidirectional videos,'' in \emph{{IEEE} International Conference on
  Multimedia and Expo}.\hskip 1em plus 0.5em minus 0.4em\relax IEEE, 2017, pp.
  799--804.

\bibitem{liu2018rate}
Y.~Liu, L.~Yang, M.~Xu, and Z.~Wang, ``Rate control schemes for panoramic video
  coding,'' \emph{Journal of Visual Communication and Image Representation},
  vol.~53, pp. 76--85, 2018.

\bibitem{li2017spherical}
Y.~Li, J.~Xu, and Z.~Chen, ``Spherical domain rate-distortion optimization for
  360-degree video coding,'' in \emph{{IEEE} International Conference on
  Multimedia and Expo}.\hskip 1em plus 0.5em minus 0.4em\relax IEEE, 2017, pp.
  709--714.

\bibitem{boyce2017spherical}
J.~Boyce and Q.~Xu, ``Spherical rotation orientation indication for {HEVC} and
  {JEM} coding of 360 degree video,'' in \emph{Applications of Digital Image
  Processing}, vol. 10396.\hskip 1em plus 0.5em minus 0.4em\relax International
  Society for Optics and Photonics, 2017, p. 103960I.

\bibitem{su2018learning}
Y.-C. Su and K.~Grauman, ``Learning compressible 360{\dg} video isomers,'' in
  \emph{{IEEE} Conference on Computer Vision and Pattern Recognition}.\hskip
  1em plus 0.5em minus 0.4em\relax IEEE, 2018, pp. 7824--7833.

\bibitem{li2018learning}
M.~Li, W.~Zuo, S.~Gu, D.~Zhao, and D.~Zhang, ``Learning convolutional networks
  for content-weighted image compression,'' in \emph{{IEEE} Conference on
  Computer Vision and Pattern Recognition}, 2018, pp. 3214--3223.

\bibitem{santurkar2018generative}
S.~Santurkar, D.~Budden, and N.~Shavit, ``Generative compression,'' in
  \emph{Picture Coding Symposium}.\hskip 1em plus 0.5em minus 0.4em\relax IEEE,
  2018, pp. 258--262.

\bibitem{lin2019efficient}
J.-L. Lin, Y.-H. Lee, C.-H. Shih, S.-Y. Lin, H.-C. Lin, S.-K. Chang, P.~Wang,
  L.~Liu, and C.-C. Ju, ``Efficient projection and coding tools for 360{\dg}
  video,'' \emph{{IEEE} Journal on Emerging and Selected Topics in Circuits and
  Systems}, vol.~9, no.~1, pp. 84--97, 2019.

\bibitem{he2018content}
Y.~He, X.~Xiu, P.~Hanhart, Y.~Ye, F.~Duanmu, and Y.~Wang, ``Content-adaptive
  360-degree video coding using hybrid cubemap projection,'' in \emph{Picture
  Coding Symposium}.\hskip 1em plus 0.5em minus 0.4em\relax IEEE, 2018, pp.
  313--317.

\bibitem{hanhart2019360}
P.~Hanhart, X.~Xiu, Y.~He, and Y.~Ye, ``360{\dg} video coding based on
  projection format adaptation and spherical neighboring relationship,''
  \emph{{IEEE} Journal on Emerging and Selected Topics in Circuits and
  Systems}, vol.~9, no.~1, pp. 71--83, 2019.

\bibitem{li2016novel}
J.~Li, Z.~Wen, S.~Li, Y.~Zhao, B.~Guo, and J.~Wen, ``Novel tile segmentation
  scheme for omnidirectional video,'' in \emph{{IEEE} International Conference
  on Image Processing}.\hskip 1em plus 0.5em minus 0.4em\relax IEEE, 2016, pp.
  370--374.

\bibitem{chengjia2018octagonal}
C.~Wu, H.~Zhao, and X.~Shang, ``Octagonal mapping scheme for panoramic video
  encoding,'' \emph{{IEEE} Transactions on Circuits and Systems for Video
  Technology}, vol.~28, no.~9, pp. 2402--2406, 2018.

\bibitem{sreedhar2016viewport}
K.~K. Sreedhar, A.~Aminlou, M.~M. Hannuksela, and M.~Gabbouj,
  ``Viewport-adaptive encoding and streaming of 360-degree video for virtual
  reality applications,'' in \emph{{IEEE} International Symposium on
  Multimedia}.\hskip 1em plus 0.5em minus 0.4em\relax IEEE, 2016, pp. 583--586.

\bibitem{taghavinasrabadi2017adaptive}
A.~T. Nasrabadi, A.~Mahzari, J.~D. Beshay, and R.~Prakash, ``Adaptive
  360-degree video streaming using layered video coding,'' in \emph{{IEEE}
  Virtual Reality}.\hskip 1em plus 0.5em minus 0.4em\relax IEEE, 2017, pp.
  347--348.

\bibitem{ozcinar2017viewport}
C.~Ozcinar, A.~De~Abreu, and A.~Smolic, ``Viewport-aware adaptive 360 video
  streaming using tiles for virtual reality,'' in \emph{{IEEE} International
  Conference on Image Processing}.\hskip 1em plus 0.5em minus 0.4em\relax IEEE,
  2017, pp. 2174--2178.

\bibitem{gopro}
A.~Abbas and B.~Adsumilli, ``New {GoPro} test sequences for virtual reality
  video coding,'' in \emph{ITU-T VCEG and ISO/IEC MPEG 4th Meeting}, 2016.

\bibitem{vrsequences}
{IEEE1857.9 1st Beijing}, ``{1857.9-01-N0001} output document,'' 2016.

\bibitem{xiao2012recognizing}
J.~Xiao, K.~A. Ehinger, A.~Oliva, and A.~Torralba, ``Recognizing scene
  viewpoint using panoramic place representation,'' in \emph{{IEEE} Conference
  on Computer Vision and Pattern Recognition}.\hskip 1em plus 0.5em minus
  0.4em\relax IEEE, 2012, pp. 2695--2702.

\bibitem{letinvr}
W.~Sun and R.~Guo, ``Test sequences for virtual reality video coding from
  {LetinVR},'' \emph{JVET of ITU-T SG16 WP3 and ISO/IEC JTC1/SC29/WG11}, 2016.

\bibitem{corbillon2017viewport}
X.~Corbillon, G.~Simon, A.~Devlic, and J.~Chakareski, ``Viewport-adaptive
  navigable 360-degree video delivery,'' in \emph{{IEEE} International
  Conference on Communications}.\hskip 1em plus 0.5em minus 0.4em\relax IEEE,
  2017, pp. 1--7.

\bibitem{bang2016description}
G.~Bang, G.~Lafruit, and M.~Tanimoto, ``Description of 360 3d video application
  exploration experiments on divergent multiview video,'' \emph{Tech. Rep.
  MPEG2015/M16129, ISO/IEC JTC1/SC29/WG11, Chengdu, CN}, 2016.

\bibitem{nguyen2019optimal}
D.~V. Nguyen, H.~T. Tran, A.~T. Pham, and T.~C. Thang, ``An optimal tile-based
  approach for viewport-adaptive 360-degree video streaming,'' \emph{{IEEE}
  Journal on Emerging and Selected Topics in Circuits and Systems}, vol.~9,
  no.~1, pp. 29--42, 2019.

\bibitem{sanchez2019delay}
Y.~Sanchez, G.~S. Bhullar, R.~Skupin, C.~Hellge, and T.~Schierl, ``Delay impact
  on {MPEG} {OMAF}'s tile-based viewport-dependent 360{\dg} video streaming,''
  \emph{{IEEE} Journal on Emerging and Selected Topics in Circuits and
  Systems}, vol.~9, no.~1, pp. 18--28, 2019.

\bibitem{nasrabadi2017adaptive}
A.~T. Nasrabadi, A.~Mahzari, J.~D. Beshay, and R.~Prakash, ``Adaptive
  360-degree video streaming using scalable video coding,'' in \emph{{ACM}
  International Conference on Multimedia}.\hskip 1em plus 0.5em minus
  0.4em\relax ACM, 2017, pp. 1689--1697.

\bibitem{ozcinar2017estimation}
C.~Ozcinar, A.~De~Abreu, S.~Knorr, and A.~Smolic, ``Estimation of optimal
  encoding ladders for tiled 360 {VR} video in adaptive streaming systems,'' in
  \emph{{IEEE} International Symposium on Multimedia}.\hskip 1em plus 0.5em
  minus 0.4em\relax IEEE, 2017, pp. 45--52.

\bibitem{sun2019two}
L.~Sun, F.~Duanmu, Y.~Liu, Y.~Wang, Y.~Ye, H.~Shi, and D.~Dai, ``A two-tier
  system for on-demand streaming of 360 degree video over dynamic networks,''
  \emph{{IEEE} Journal on Emerging and Selected Topics in Circuits and
  Systems}, vol.~9, no.~1, pp. 43--57, 2019.

\bibitem{ghaznavi2017comparison}
R.~Ghaznavi-Youvalari, A.~Zare, H.~Fang, A.~Aminlou, Q.~Xie, M.~M. Hannuksela,
  and M.~Gabbouj, ``Comparison of hevc coding schemes for tile-based
  viewport-adaptive streaming of omnidirectional video,'' in \emph{{IEEE}
  International Workshop on Multimedia Signal Processing}.\hskip 1em plus 0.5em
  minus 0.4em\relax IEEE, 2017, pp. 1--6.

\bibitem{lederer2017today}
S.~Lederer, ``Today's and future challenges with new forms of content like
  360{\dg}, {AR} and {VR},'' in \emph{{MPEG} Workshop: Global Media Technology
  Standards for an Immersive Age (Invited Talk)}.\hskip 1em plus 0.5em minus
  0.4em\relax MPEG, 2017.

\bibitem{zare2017virtual}
A.~Zare, A.~Aminlou, and M.~M. Hannuksela, ``Virtual reality content streaming:
  Viewport-dependent projection and tile-based techniques,'' in \emph{{IEEE}
  International Conference on Image Processing}.\hskip 1em plus 0.5em minus
  0.4em\relax IEEE, 2017, pp. 1432--1436.

\bibitem{skupin2016tile}
R.~Skupin, Y.~Sanchez, C.~Hellge, and T.~Schierl, ``Tile based {HEVC} video for
  head mounted displays,'' in \emph{{IEEE} International Symposium on
  Multimedia}.\hskip 1em plus 0.5em minus 0.4em\relax IEEE, 2016, pp. 399--400.

\bibitem{hosseini2016adaptive}
M.~Hosseini and V.~Swaminathan, ``Adaptive 360 {VR} video streaming: Divide and
  conquer,'' in \emph{{IEEE} International Symposium on Multimedia}.\hskip 1em
  plus 0.5em minus 0.4em\relax IEEE, 2016, pp. 107--110.

\bibitem{alface2012interactive}
P.~R. Alface, J.-F. Macq, and N.~Verzijp, ``Interactive omnidirectional video
  delivery: A bandwidth-effective approach,'' \emph{Bell Labs Technical
  Journal}, vol.~16, no.~4, pp. 135--147, 2012.

\bibitem{zhu2018innovative}
C.~Zhu, K.~Huang, and G.~Li, ``An innovative saliency guided roi selection
  model for panoramic images compression,'' in \emph{Data Compression
  Conference}.\hskip 1em plus 0.5em minus 0.4em\relax IEEE, 2018, pp. 436--436.

\bibitem{biswas2017towards}
S.~Biswas, S.~A. Fezza, and M.-C. Larabi, ``Towards light-compensated saliency
  prediction for omnidirectional images,'' in \emph{International Conference on
  Image Processing Theory, Tools and Applications}.\hskip 1em plus 0.5em minus
  0.4em\relax IEEE, 2017, pp. 1--6.

\bibitem{argyriou2007multi}
A.~Argyriou, T.~Evgeniou, and M.~Pontil, ``Multi-task feature learning,'' in
  \emph{Advances in Neural Information Processing Systems}, 2007, pp. 41--48.

\end{thebibliography}

%

\end{document}